# Physical aging of glasses of an organic semiconductor

Shinian Cheng[1*], Kritika Jha[2], Zijian Wang[3], Juliana B. Lugo[2], Hayley Kositzke[1], John H. Perepezko[3], Zahra Fakhraai[2], Mark D. Ediger[1]

[1]Department of Chemistry, University of Wisconsin-Madison, Madison, Wisconsin 53706, United States

[2]Department of Chemistry, University of Pennsylvania, Philadelphia, Pennsylvania 19104, United States

[3]Department of Materials Science and Engineering, University of Wisconsin-Madison, Madison, Wisconsin 53706, United States

***Corresponding Author:** chengshinian@gmail.com

**Abstract:** All glasses, including organic semiconductor glasses, are non-equilibrium materials whose properties will change with time. This "physical aging" process is poorly understood for organic semiconductors, hindering the rational design of highly durable devices. In this study, we investigated the volume and enthalpy recovery processes in both thin films and bulk glasses of N,N'-Bis(3-methylphenyl)-N,N'-diphenylbenzidine (TPD). Our results revealed that volume recovery kinetics exhibit negligible dependence on film thickness for liquid-cooled TPD films between 400 nm and 100 nm. Additionally, the volume recovery process in TPD films was strongly coupled to the enthalpy recovery observed in bulk TPD glasses during annealing near the glass transition temperature. Remarkably, TPD films prepared by physical vapor deposition at room temperature demonstrated exceptional resistance to physical aging, with an aging rate approximately one order of magnitude lower than that of their liquid-cooled counterparts. These results not only enhance our understanding of the non-equilibrium dynamics in amorphous systems but also offer valuable insights for the design of next-generation organic devices with significantly improved stability and durability.

# The table of contents entry:

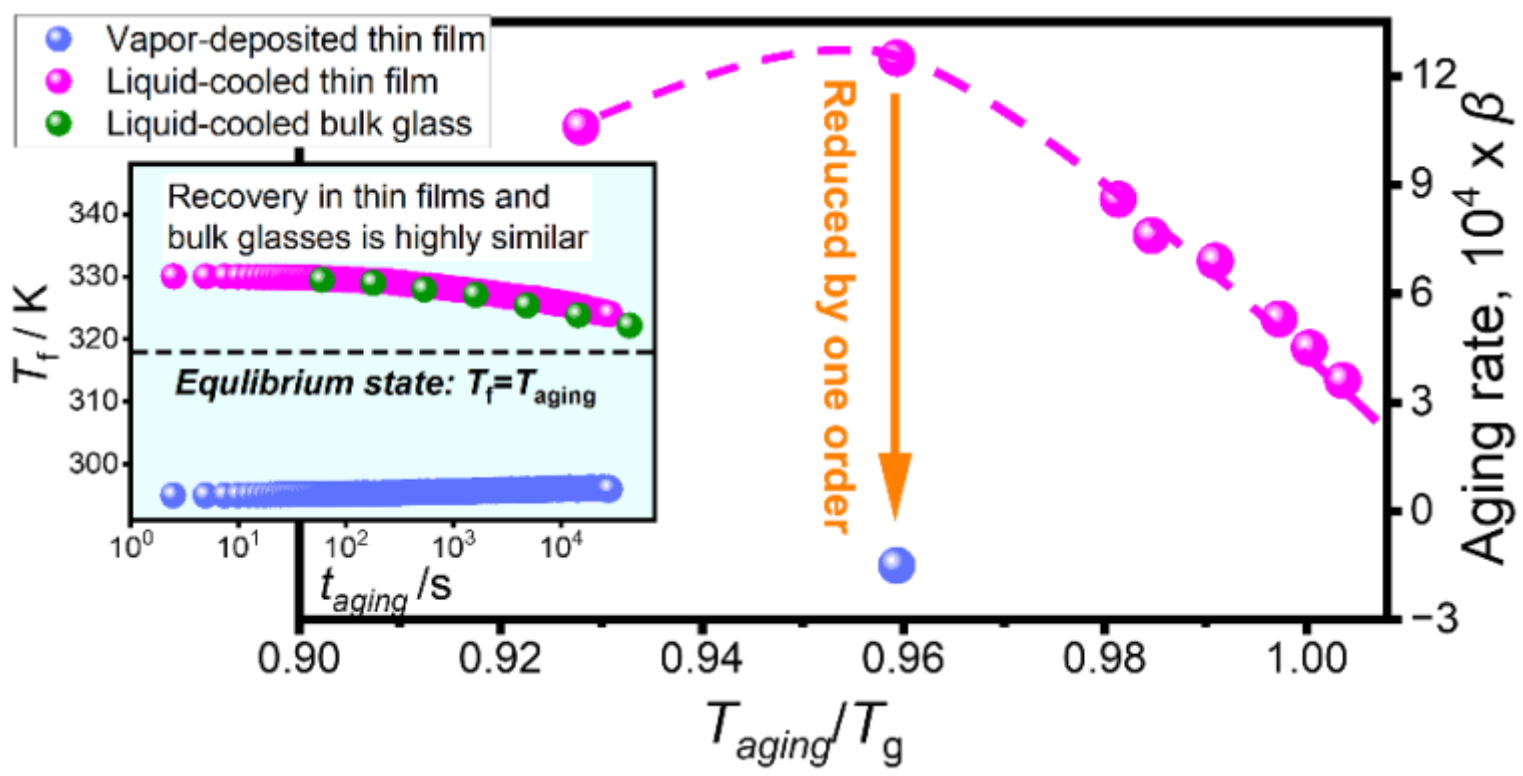

**Key findings:** Organic semiconductor films and their bulk counterparts demonstrate remarkably similar recovery kinetics. Vapor-deposited films display significantly enhanced resistance to physical aging compared to liquid-cooled films.

## Introduction

Organic semiconductors have emerged as essential materials in the realm of organic electronic devices, offering a unique combination of flexibility, lightweight properties, and ease of fabrication[1-5]. For applications in organic electronics that require macroscopic homogeneity and smooth surfaces, such as organic light-emitting diodes (OLEDs), organic semiconductor glasses offer better performance than that can be achieved with crystals and play a crucial role in advancing smart and wearable devices, such as flexible displays[3, 6, 7]. However, due to their inherently non-equilibrium nature, glassy materials undergo a spontaneous process known as structural recovery or physical aging, evolving toward equilibrium over time[8-13].

Physical aging is caused by molecular motion and can lead to important changes in physical properties, such as volume[14, 15], enthalpy[16, 17], and modulus[10, 18]. For organic semiconductor glasses, small changes in molecular orientations and the positions of nearest neighbors allow the system to find more efficient and lower energy packing arrangements; these molecular motions lower the free energy and ultimately allow the system to achieve equilibrium. The molecular motions that enable physical aging can also lead to crystallization[19], which is a major failure mode for OLEDs[20, 21]. Therefore, a deeper understanding of physical aging in organic semiconductor glasses is essential for designing more durable electronic devices. Despite its significance, limited research has been conducted on the physical aging of organic semiconductors[22, 23], particularly in very thin films, which are widely used in practical applications.

Volume and enthalpy recovery measurements are two primary experimental approaches commonly utilized to investigate the physical aging behavior of amorphous materials[10, 15]. While both measurements reflect how these materials gradually relax toward thermodynamic equilibrium over time, they can offer complementary perspectives on different aspects of the aging process. Volume recovery provides insight into how molecular packing evolves over time, which is crucial for understanding changes in the macroscopic dimensions and density of the materials. Enthalpy recovery illustrates the way a material releases stored internal energy as it moves towards a lower energy state, revealing changes in the position of the system on the potential energy landscape. Despite important investigations[24], there is no generic relationship between volume and enthalpy recovery kinetics[15, 25], complicating the characterization of materials during physical aging.

Therefore, it is useful to study both volume and enthalpy recovery processes in order to obtain a more comprehensive understanding of physical aging in organic semiconductors.

Studies in recent decades have also shown that sample size and geometry can have a significant impact on physical aging in amorphous materials[26-28]. There has not been a systematic study of physical aging in the ~ 100 nm films of organic semiconductors which are relevant for devices. There is substantial evidence that the physical aging behavior of thin films can differ significantly from that observed in bulk[27]. The effect of film thickness on physical aging kinetics and rate is sufficiently complex that one cannot predict for organic semiconductors the extent to which film thickness will alter physical aging.

A further relevant aspect of physical aging is the influence of the initial state of the glass on the aging kinetics. Recent work has shown that glasses prepared by physical vapor deposition (PVD) can have significantly higher density[29-31] and lower enthalpy[32-34] than liquid-cooled glasses. This comparison is particularly relevant in the context of OLEDs since current commercial devices are prepared by PVD, but there is considerable interest in switching to solution processing (which produces glasses similar to those prepared by liquid-cooling). Based on previous reports[35-39], it is anticipated that glasses prepared by PVD will age much more slowly than those prepared by liquid-cooling or solution processing, but these differences have not been systematically quantified.

In this work, we investigated the physical aging behavior of N,N'-Bis(3-methylphenyl)-N,N'-diphenylbenzidine (TPD) glasses, a material widely utilized in thin films to produce OLED devices[40, 41] and as a model system for studying the impact of PVD on glass properties[23, 36, 42, 43]. Our investigation included measurements of volume and enthalpy relaxation in TPD thin films and bulk samples. For thin films, we have compared the aging behavior of liquid-cooled glasses to that of PVD glasses. Spectroscopic ellipsometry was utilized to monitor thickness changes in TPD thin films, providing a direct measurement of volume recovery. Conventional and Flash differential scanning calorimetry was employed to measure the enthalpy recovery of slow-cooled and hyper-quenched TPD bulk glasses. We found that liquid-cooled glasses of TPD physically aged to equilibrium in the time window of our experiments (~30000 s) when annealed near the glass transition temperature. At lower temperatures, substantial physical aging occurs but equilibrium is not attained within our time window. At a given annealing temperature, the timescales required for physical aging were consistent between the volume recovery of thin films and the enthalpy recovery of bulk systems, demonstrating a strong coupling between these two processes. Notably, volumetric aging was almost identical for 400 nm and 100 nm liquid-cooled glasses of TPD, indicating negligible thickness effects on physical aging kinetics in this thickness range. Furthermore, we report that TPD glass films prepared by PVD at room temperature exhibited exceptional resistance to aging compared to liquid-cooled TPD films; for annealing at 318 K, the physical aging rate for vapor-deposited TPD films was suppressed by almost one order of magnitude compared to liquid-cooled glasses.

## Materials and Methods

***TPD film preparation*:** TPD (99% purity) was purchased from Sigma-Aldrich and used without further purification. Vapor-deposited glass films of TPD were prepared on Si substrates at 298 K in a high-vacuum chamber with a base pressure ~$10^{-6}$ Torr. The film thickness and deposition rate (0.20±0.02 nm/s) were monitored using a quartz crystal microbalance during deposition. By heating the vapor-deposited TPD films to 355 K, TPD films in the supercooled liquid state were obtained. Subsequently, these films were cooled to prepare the liquid-cooled glasses used in the aging experiments.

***Spectroscopic ellipsometry (SE)*:** The optical properties of the TPD films were measured using a spectroscopic ellipsometer (J.A. Woollam M-2000U) with a custom-built temperature control and translation stage. The temperature and time dependent measurements were performed at a fixed incident angle of 70°. The collected optical quantities in the wavelength range from 550 nm to 1000 nm were fitted to a three-layer model consisting of the silicon substrate layer, 2 nm thick native oxide layer, and the transparent sample layer (TPD film). The isotropic Cauchy model ($n = A + \frac{B}{\lambda^2} + \frac{C}{\lambda^4}$) was applied to determine the thickness of liquid-cooled TPD film. For vapor-deposited TPD films, we utilized the anisotropic Cauchy model ($n_{xy} = A_{xy} + \frac{B}{\lambda^2} + \frac{C}{\lambda^4}$; $n_z = A_z + \frac{B}{\lambda^2} + \frac{C}{\lambda^4}$); here the birefringence ($\Delta n = n_z - n_{xy}$) characterizes anisotropic packing. The representative Cauchy model fitting parameters and mean-squared-errors (MSE) at $T_{\text{aging}}$=318K were displayed as a function of aging time in Figs.S2-5 in Supporting Information (SI).

***Differential Scanning Calorimetry (DSC)***: The thermal properties of bulk TPD samples were measured using a TA Q2000 Differential Scanning Calorimeter. The TPD samples (≈6 mg) were carefully mounted using T-zero Aluminum pans (TA Instruments) sealed with hermetic lids. The aging experiments were performed under 50 mL/min $N_2$ purge.

***Flash Differential Scanning Calorimetry (FDSC)***: The thermal properties of bulk TPD were also studied using a high-rate differential scanning calorimeter (Flash DSC 2+, Mettler Toledo). During measurement, argon gas flow was applied to protect the sample from oxidation. The sample mass was estimated based on the heat capacity change at $T_g$, and found to be approximately 500 ng. Based on the sample mass and the chip sensor area, the average sample thickness was estimated to be around 10 μm. Prior to mounting the sample, the Flash DSC chip sensor was conditioned and corrected using the procedure provided by Mettler Toledo. Post-measurement temperature calibration was performed using tin as the reference substance.

***Aging experiments***: Volume recovery experiments on TPD glasses were performed using SE. For aging experiments on liquid-cooled TPD glasses, the temperature profile of one complete cycle is displayed in Fig. S1a in Supporting information (SI). To ensure reproducibility, two complete cycles were performed for each aging temperature. For aging experiments on vapor-deposited TPD glasses, after deposition, TPD films were immediately moved to the ellipsometry stage and heated up to 318K. The temperature profile is shown in Fig. 3a below. For all volume recovery experiments on TPD glasses, at least three different samples were measured for a specific aging temperature and both cooling and heating rates were 5K/min.

Enthalpy recovery experiments on bulk TPD were performed using DSC and FDSC. The temperature profiles of one complete cycle are displayed in Figs. S1b and S1c, for DSC and Flash DSC respectively, and two complete cycles were performed for each aging temperature. The cooling and heating rates were 5K/min for DSC, and 1000K/s for FDSC, respectively. To ensure reproducibility, the enthalpy recovery experiments were performed on at least three different samples for a specific aging temperature.

## Results and Discussion

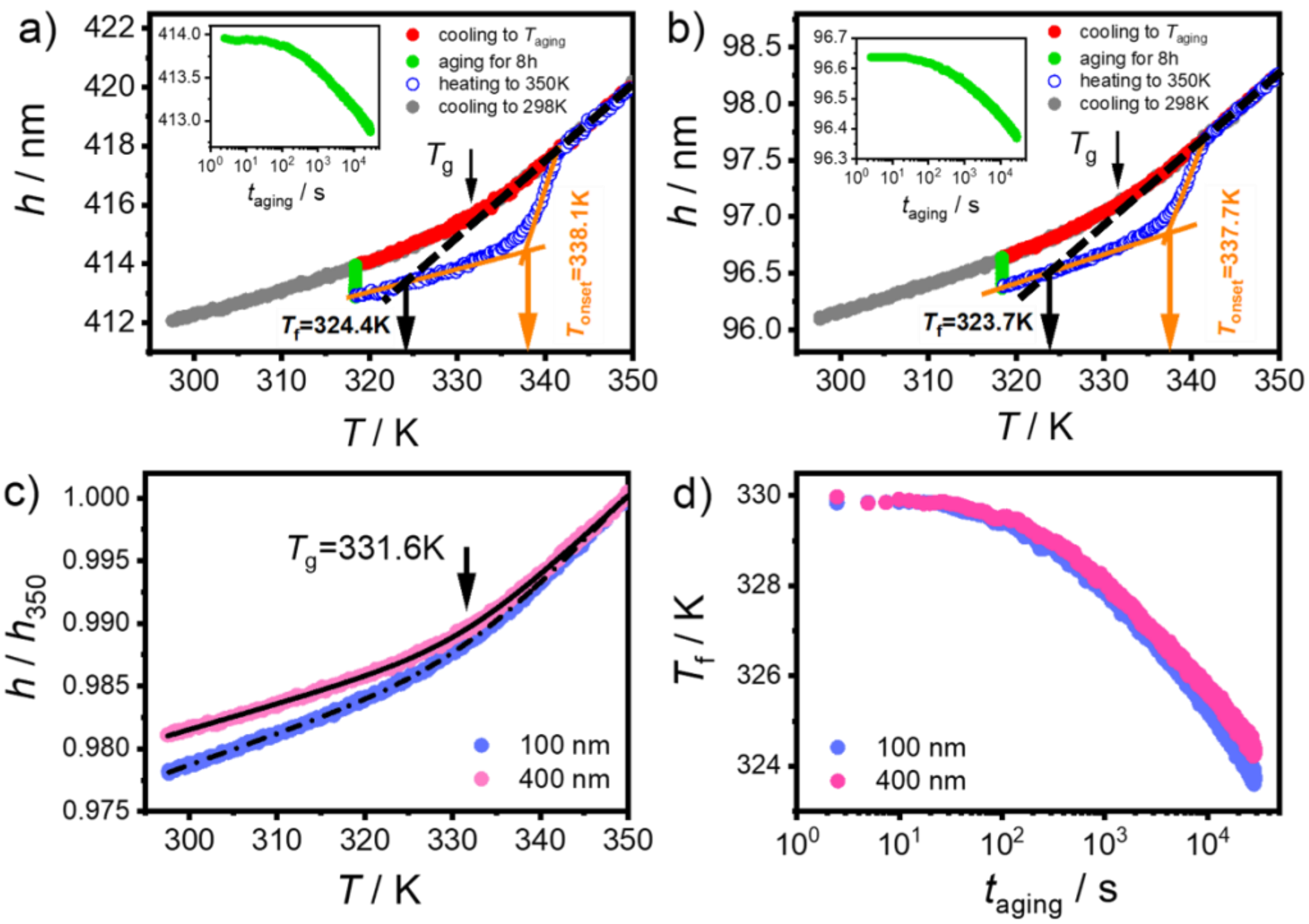

**Fig. 1:** Volume evolution of TPD glasses and the supercooled liquid. **a)** Thickness as a function of temperature for a 400 nm liquid-cooled TPD glass during an aging experiment at $T_{aging}$=318 K. **b)** Thickness as a function of temperature for a 100 nm TPD film during an aging experiment at $T_{aging}$=318 K. The dashed lines in a) and b) denote the extrapolation of equilibrium thickness of the TPD liquid. **Insets in a) and b):** Evolution of film thickness with aging time when annealing at 318 K for 8 h. **c)** Normalized film thickness as a function of temperature for 100 nm and 400 nm TPD films during cooling rate at 5 K/min. $h_{350}$ is the film thickness at $T$=350 K. The black lines present the fits to Eq. 1, for determining the glass transition temperatures (which are the same for 100 nm and 400 nm films, within experimental error) and the thermal expansion coefficients. **d)** Comparison of the time-dependent fictive temperature of 100 nm and 400 nm TPD glasses at $T_{aging}$=318 K.

Physical aging of liquid-cooled TPD glasses leads to increased density and enhanced kinetic stability. Figs. 1a and 1b show the thickness evolution with temperature before and after aging at $T_{aging}$=318 K for 400 nm and 100 nm TPD films, respectively. TPD films, initially in the supercooled liquid state, were first cooled to $T_{aging}$ (solid red symbols), then aged for 8 h at $T_{aging}$ (solid green symbols), and then heated to 350 K (open blue symbols). The fully transformed films were subsequently cooled down to room temperature (solid gray symbols). As displayed in Figs. 1a and 1b, the thickness of TPD glassy films after aging is significantly reduced, compared to that before aging. The evolution of film thickness during aging is displayed in the insets of Figs. 1a and 1b, where TPD film thickness continuously decreases with aging time $t_{aging}$. TPD glasses become denser during aging, which is similar to the observations in other kinds of glasses[27, 44], and consistent with the non-equilibrium nature of glasses. As can be seen from Figs. 1a and 1b, the onset temperature $T_{onset}$, at which the aged glasses start to transform into supercooled liquids upon heating, is significantly higher than the glass transition temperature $T_g$ obtained upon cooling, demonstrating the enhanced kinetic stability of the aged TPD glasses. As expected, the TPD film thickness in the supercooled liquid state is the same before and after temperature cycling and aging, as demonstrated in Figs. 1a and 1b. (Fig. S6 presents the refractive index $n$ at 632.8 nm as a function of temperature during an aging experiment at $T_{aging}$=318K for 400 nm and 100 nm TPD glasses and supercooled liquid. The refractive index data is consistent with the thickness data shown in Figs. 1a and 1b.)

From the data in Fig. 1, we observe that the initial thickness of the TPD film has a negligible effect on the volume recovery kinetics. To compare the physical aging behavior of 400 nm and 100 nm TPD films, the fictive temperature $T_f$, quantifying the thermodynamic state of aged glasses, was applied. At $T_f$, the extrapolated volume/thickness of the supercooled liquid would be equal to that of the glass. As shown in Figs. 1a and 1b, the determined $T_f$ values are almost the same for 400 nm ($T_f$=324.4 K) and 100 nm ($T_f$=323.7 K) TPD films after aging at 318K for 8h. Furthermore, for both TPD films aged for 8h, almost the same $T_{onset}$ (338 K) was found.

The thinner TPD films exhibit slightly higher thermal expansion coefficients but show the same glass transition temperature as thicker films. Fig. 1c shows the normalized film thickness during cooling for 400 nm and 100 nm TPD films. As shown, the 100 nm TPD film shows a greater thermal expansion coefficient than the 400 nm film, particularly in the glassy state. Quantitatively, the following equation proposed by Dalnoki-Veress et al.[45] was applied to determine the thermal expansion coefficients and glass transition temperature of the studied films:

$$h(T)/h_{350} = w * \frac{M-G}{2} * ln\left[\cosh\left(\frac{T-T_g}{w}\right)\right] + (T - T_g) * \frac{M+G}{2} + c \qquad (1)$$

In this equation, $M$ and $G$ are the thermal expansion coefficients of the supercooled liquid and glassy regimes, $w$ is the width of the glass transition, and $c$ is the normalized film thickness at $T_g$. The black lines in Fig. 1c represent the fitting results with the parameters: $M = (6.8 \pm 0.1) \times 10^{-4}$ $K^{-1}$, $G = (2.0 \pm 0.1) \times 10^{-4}$ $K^{-1}$, and $T_g = 331.9 \pm 0.3$ K for 400 nm TPD film; $M = (7.4 \pm 0.1)$

$\times 10^{-4}$ K$^{-1}$, $G = (2.5 \pm 0.1) \times 10^{-4}$ K$^{-1}$, and $T_g$= 331.6 ± 0.3 K for 100 nm TPD film. The determined $T_g$ values are identical within error, indicating that the thickness has no significant influence on the glass transition dynamics of studied TPD films. The thermal expansion coefficients are 8% and 20% higher in 100 nm films, compared to those in 400 nm films, in the supercooled liquid and glassy states, respectively. The expansion coefficients and glass transition temperatures obtained here are in reasonable agreement with those reported by Zhang et al.[46] for 115 nm TPD films using a 10 K/min cooling rate, where $M = (6.6 \pm 0.3) \times 10^{-4}$ K$^{-1}$, $G = (2.1 \pm 0.03) \times 10^{-4}$ K$^{-1}$, and $T_g$=330 ± 1 K.

Moreover, the evolution of $T_f$ with aging time $t_{aging}$ exhibited nearly the same trend between 400 nm and 100 nm TPD films. With the obtained expansion coefficients at glassy and supercooled liquid states, we determined $T_f$ of the aged 400 nm and 100 nm TPD films at 318K and these are plotted as a function of aging time in Fig. 1d. These results demonstrate that 400 nm and 100 nm TPD films present essentially the same volume recovery kinetics when annealing at 318 K. Boucher *et al.*[47] have connected an acceleration of physical aging kinetics in very thin polystyrene films to a depression of $T_g$. From this perspective, the highly similar aging behavior observed in 400 nm and 100 nm TPD films is consistent with their identical $T_g$ values. Previous study has shown that significant reduction of $T_g$ starts when thickness is less than 30 nm for TPD films[46]. Fig. S7 illustrates the method applied to determine $T_f$ of the TPD films.

Lowering aging temperature elongates the timescale for reaching the equilibrium state. Figs. 2a displays the evolution of normalized film thickness with aging time for 400 nm TPD films at various aging temperatures. When annealing at higher temperatures near $T_g$ (*i.e.*, $T_{aging}$=333, 332, 331 K), film thickness levels off in our experimental window (as indicated by dashed lines), while detectable decreases in film thickness remain at the longest aging times for lower aging temperatures. Similar features can be seen from the plots of $T_f$ versus $t_{aging}$ in Fig. 2b. For physical aging experiments at the lower temperatures, the final $T_f$ values after aging for 8h are significantly higher than the corresponding $T_{aging}$. This indicates that the aged glasses at lower temperatures are still considerably less dense than the corresponding equilibrium state and that much longer times would be required to age to equilibrium.

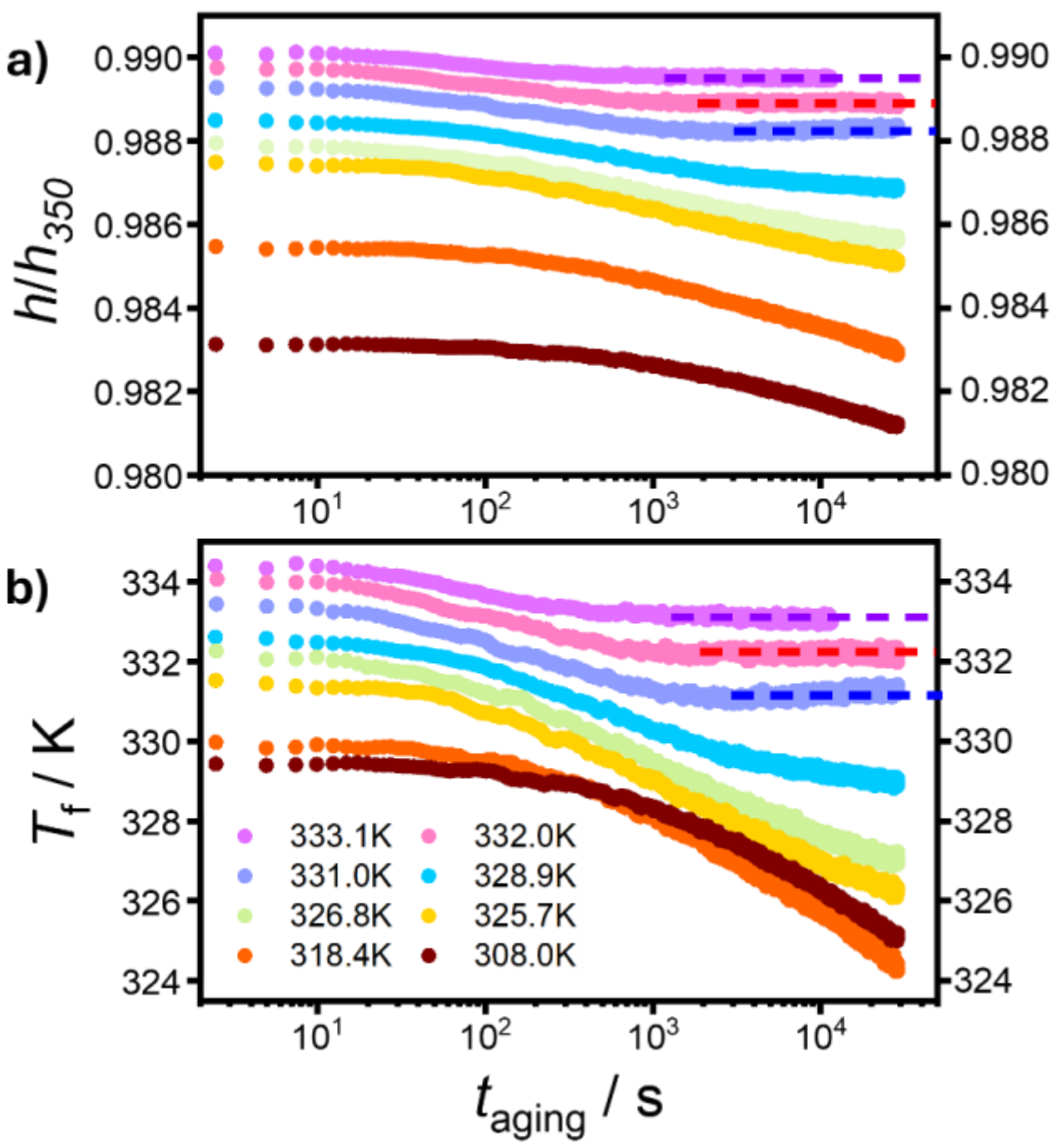

**Fig.2:** The evolution of normalized film thickness (a) and fictive temperature (b) with aging time for 400 nm TPD films, at different aging temperatures as indicated in the figure. The dashed lines are drawn at constant $h/h_0$ (a) and constant $T_f$ (b) to indicate the behavior at long aging times. In this plot, the room temperature thicknesses of the films range from 390 nm to 410 nm, and $h_{350}$ is the absolute thickness of each film at $T$=350 K.

Remarkably, TPD films prepared by vapor deposition at room temperature exhibited exceptional resistance to physical aging. Studying the physical aging properties of vapor-deposited organic semiconductors has practical implications since they are widely used in devices. Fig. 3a presents the experimental profile of physical aging measurements on vapor-deposited and liquid-cooled TPD films. The aging experiments were performed at $T_{aging}$=318 K with $t_{aging}$=8 h for both vapor-deposited and liquid-cooled TPD glasses. Fig. 3b shows the temperature-dependent thickness of TPD films before and after annealing at 318 K for 8 h. Consistent with previous reports, TPD films deposited at room temperature exhibit significantly increased density (or reduced thickness)[36]. The evolution of $T_f$ with aging time for both vapor-deposited and liquid-cooled TPD films during aging is displayed in Fig. 3c. While significant reduction in $T_f$ (from 330K to 324K) is observed in liquid-cooled TPD films, the $T_f$ of vapor-deposited TPD remains almost unchanged in the experimental time window, highlighting the significantly reduced volume recovery kinetics. This observation is in accord with the formation of "ultra-stable" TPD glasses[36] where the structural relaxation is suppressed significantly. Accordingly, Fig. S8 displays the thickness evolution of both PVD and liquid-cooled TPD glass with aging time at $T_{aging}$=318K. While significant decrease in thickness is observed for the liquid-cooled glass, we see a slight thickness increase (~0.05% from 391.7 nm to 391.9 nm) during aging for vapor-deposited TPD thin films. For the PVD glass, an increase in thickness (decrease in density) is expected during aging, given that the PVD glass starts at higher density than the equilibrium supercooled liquid, as clearly illustrated in Fig. 3c.

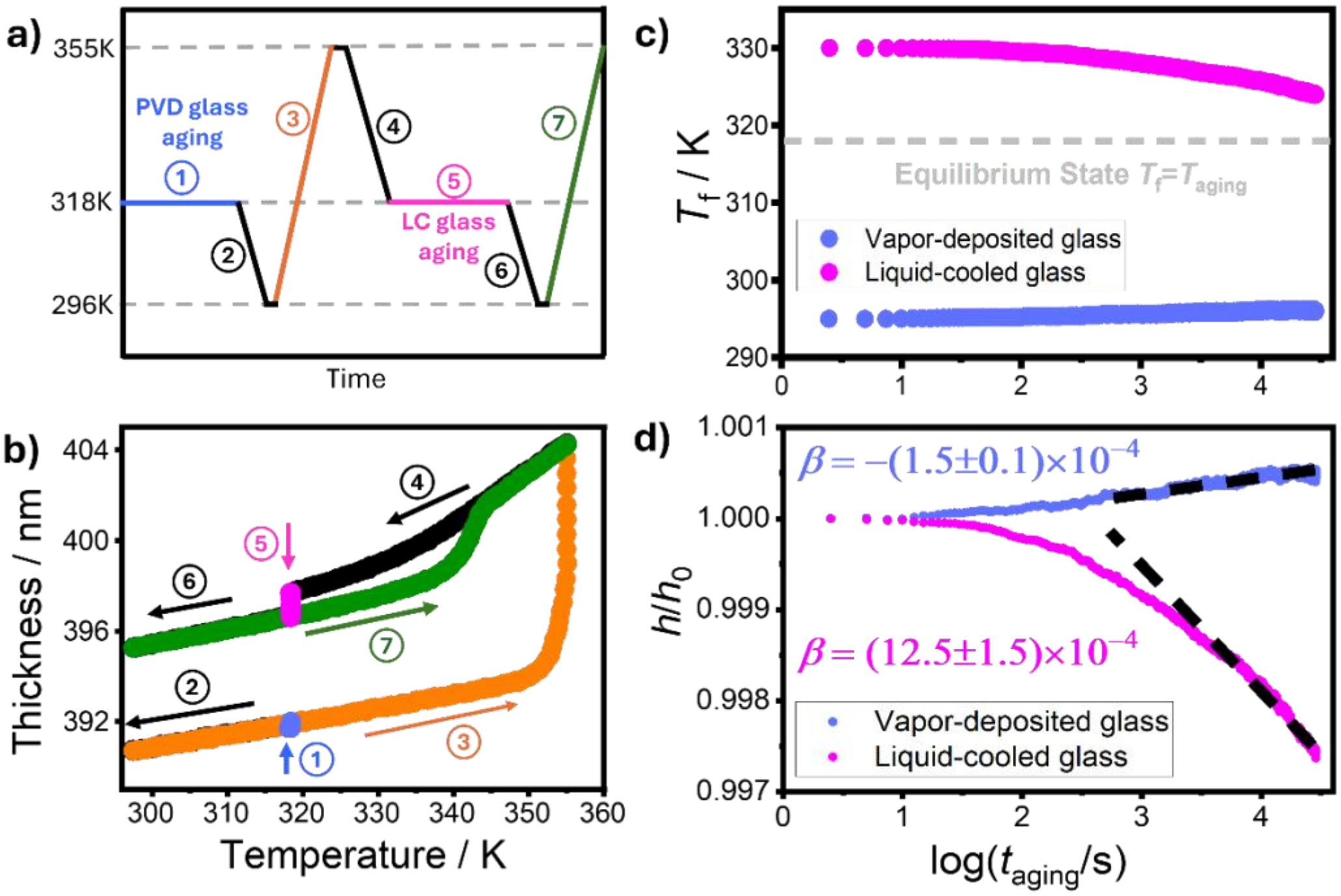

**Fig. 3.** Comparison of physical aging for vapor-deposited and liquid-cooled glasses of TPD. **a)** The temperature profile used to compare aging in vapor-deposited and liquid-cooled glasses; **b)** Thickness as a function of temperature for vapor-deposited and liquid-cooled TPD glasses; **c)** The evolution of $T_f$ with aging time at $T_{aging}$=318 K for vapor-deposited (blue) and liquid-cooled (pink) TPD glasses. The gray dashed line shows the extrapolated equilibrium state with $T_f$=$T_{aging}$=318 K. **d)** Normalized film thickness is plotted as a function of aging time. The dashed black lines present the linear fits used to obtain physical aging rates of vapor-deposited and liquid-cooled TPD glasses. Here $h_0$ is the film thickness at $t_{aging}$=0.

It is useful to quantify the physical aging rate for volume recovery in these TPD glasses, in order to make comparisons with other types of glasses. To evaluate the volume recovery rate of TPD films at different aging temperatures, this equation from the polymer glass research area was utilized[27]:

$$\beta = -\frac{d(\frac{h}{h_0})}{d(\log t_{aging})} \tag{2}$$

In our analysis, $h_0$ was set to be the film thickness at $t_{aging}$=0 s. Fig. 3d and Fig. S9 show the $h/h_0$ *vs* log $t_{aging}$ plots of TPD glasses, and the black dashed lines illustrate the fits used to determine the volume recovery rates. The obtained $\beta$ values for liquid-cooled and vapor-deposited TPD glasses are plotted as a function of $T_{aging}/T_g$ in Fig. 4. In this figure, we also plot the volume recovery rate for liquid-cooled polystyrene glasses reported by Pye *et al.*[27]. The polystyrene data shown here was obtained for a film with thickness of 2430 nm, which was shown to exhibit bulk aging behavior. Polystyrene is an appropriate system for comparison as it has been extensively studied and has aging behavior typical of polymer glasses. In calculating the physical aging rate, we chose the $t_{aging}$ window where significant physical aging occurs. The error bars in Fig. 4 indicate the variation in the aging rate based on the selection of different aging time intervals.

In Fig. 4, a common pattern is observed for the volume recovery rate data for liquid-cooled TPD and polystyrene glasses. As the aging temperature increases from low temperature, the volume recovery rate increases due to increased molecular mobility. In the high temperature regime, the volume recovery rate decreases with increasing aging temperature due to a diminished thermodynamic driving force for aging. At intermediate temperatures, these two effects are balanced to produce the highest rate of volume recovery. While liquid-cooled glasses of both polystyrene and TPD exhibit this qualitative pattern, the highest rate of volume recovery occurs considerably closer to $T_g$ for TPD. The highest volume recovery rate obtained here for liquid-cooled TPD films is $(12.5 \pm 1.5) \times 10^{-4}$, in comparison to the value for polystyrene of $(10.2 \pm 1.2) \times 10^{-4}$.

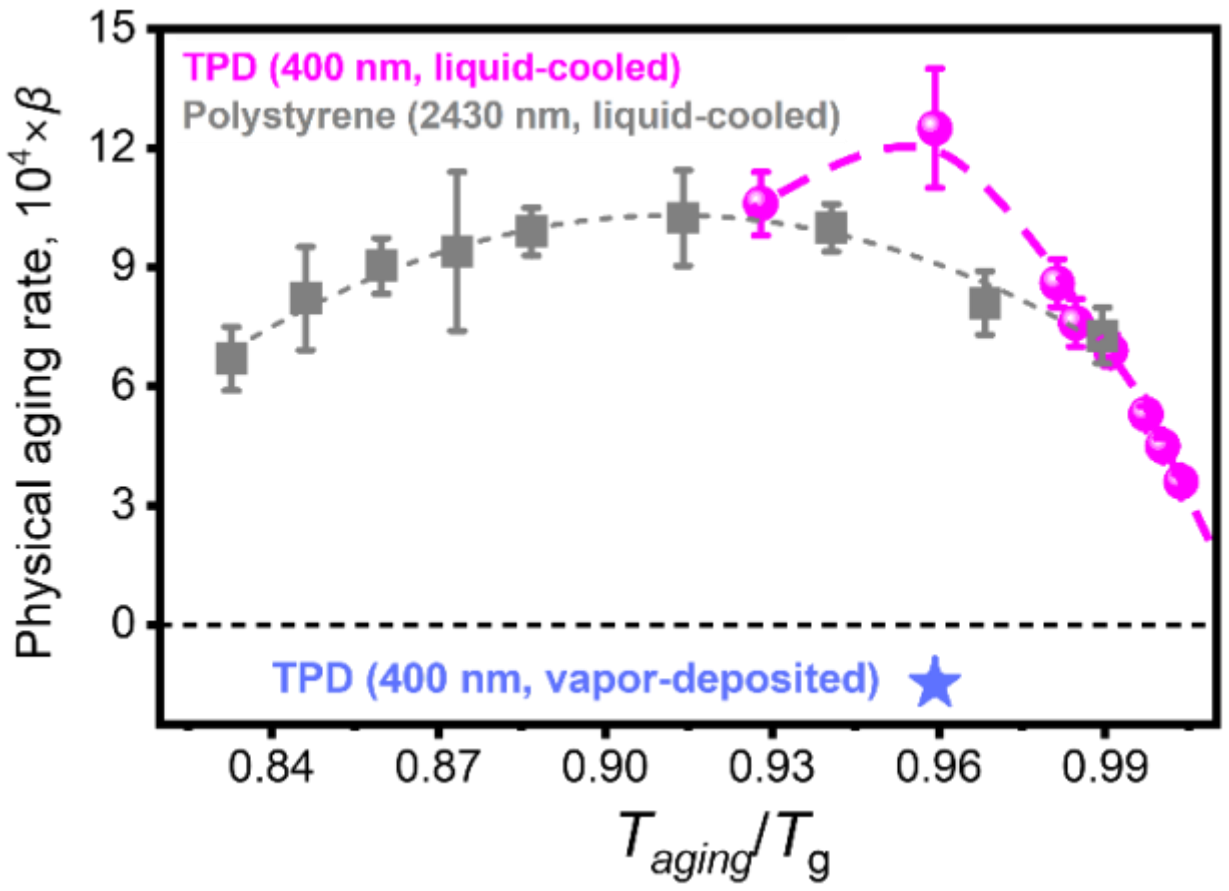

**Fig. 4.** The physical aging rate of liquid-cooled (pink) and vapor-deposited (blue) TPD glasses is plotted as a function of $T_{aging}/T_g$ and compared to data for polystyrene films, taken from ref.[27]. The dashed lines are guides for the eyes.

More interestingly, the volume recovery rate of vapor-deposited TPD glasses is quite close to zero and its absolute value is approximately one order of magnitude lower than that of the corresponding liquid-cooled glass. A negative recovery rate again confirms that the vapor-deposited TPD glass starts at higher density than the equilibrium supercooled liquid. To our knowledge, this is the first direct quantification of the physical aging rate of vapor-deposited thin glassy films. We note that volume recovery rates much closer to zero than the present work would likely be obtained if TPD were vapor-deposited at a lower temperature. The present samples were deposited using a substrate temperature of 0.90 $T_g$. Moratalla et al. showed that deposition at 0.85 $T_g$ prepares a glass with even higher kinetic stability[48], as quantified by the onset temperature. For this study, we choose a substrate temperature of 0.90 $T_g$ (298 K), because the commercial production of OLEDs is thought to utilize room temperature deposition. The comparison shown in Figures 3 (c and d) is also likely relevant for OLEDs prepared by solution processing. Recent work has shown the spin-coated and liquid-cooled glasses of organic semiconductors have similar kinetic stabilities.[49] Thus, it is likely that OLEDs prepared by solution processing will physically age more rapidly than those prepared by PVD.

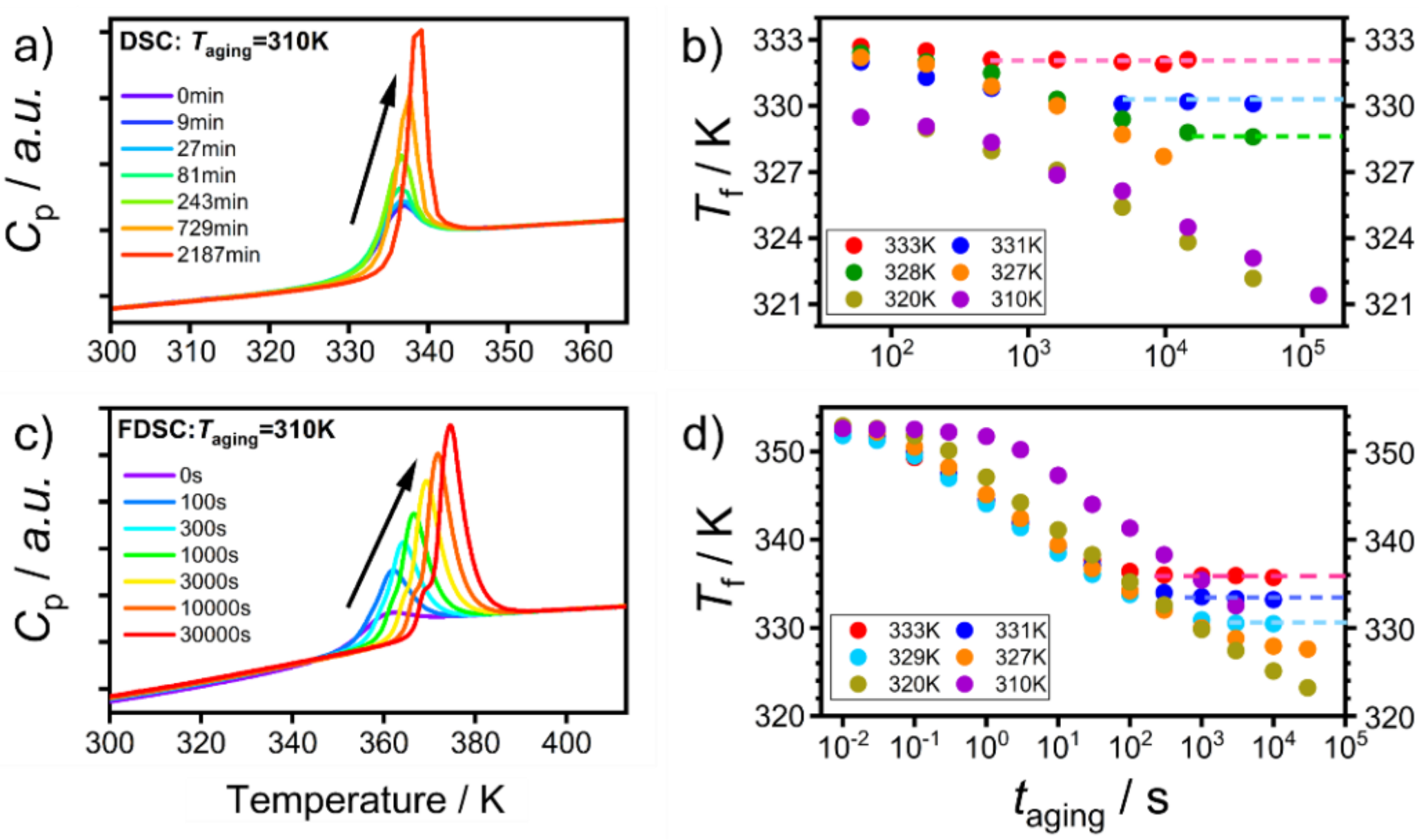

**Fig. 5:** Representative heat capacity data as a function of temperature for bulk TPD glasses aged at 310 K for various aging times, using **a)** DSC and **c)** FDSC. The black arrows in (a) and (c) indicate the evolution of the endothermic peak with increasing aging time. $T_f$ as a function of aging time at various aging temperatures for **b)** DSC and **d)** FDSC. The dashed lines denote a constant $T_f$ regime, indicating that equilibrium is attained at long aging times. Note that panels b and d show different numbers of decades on the x-axis.

To obtain a more complete picture of physical aging in TPD glasses, the enthalpy recovery of slow-cooled and hyper-quenched TPD bulk glasses was studied by using DSC and FDSC, respectively. Figs 5a and 5c show representative DSC and FDSC thermograms after annealing at 310 K over a wide aging time range. These thermograms display the typical effects of physical aging on the temperature-dependent heat capacity, $C_p$. As an effect of physical aging, an endothermic peak develops in the glass transition region, with magnitude and maximum peak temperature rising with increasing the aging time. According to the relationship between enthalpy $H$ and heat capacity $C_p$ (*i.e.*, $C_p = dH/dT$ ) at constant pressure, the peak area between the aged and unaged ($t_{aging}$=0 s) scans is equal to the enthalpy relaxed during physical aging. After the annealed glasses are fully transformed into supercooled liquids, their heat capacities become equivalent, as required given the equilibrium nature of the supercooled liquids. These enthalpy recovery results are consistent with the volume recovery results above, as they show higher kinetic stability as aging time increases; they extend the volume recovery results by showing that TPD glasses approach lower energy states during annealing. The DSC and FDSC curves of bulk TPD glasses annealed at other temperatures are displayed in Figs. S10 and S11.

To quantitatively evaluate the enthalpy recovery data, we again make use of the fictive temperature ($T_f$), now defined as the temperature where the extrapolated equilibrium enthalpy equals that of the glasses. Figs. 5b and 5d show the determined $T_f$ as a function of $t_{aging}$ at various aging temperatures for DSC and FDSC measurements, respectively. We observe, as $t_{aging}$ increases, the spontaneous evolution of the glass configuration toward lower free enthalpy, shown as a reduction of $T_f$, until reaching the equilibrium state at $T_f$=$T_{aging}$. As indicated by the dashed lines in Figs. 5b and 5d, at aging temperatures near $T_g$, $T_f$ evolves to $T_{aging}$ within the error of the measurements, indicating that the equilibrium state is reached in the experimental timescale. At lower aging temperatures, a detectable decrease in $T_f$ is still observed even at the longest aging times. These observations in enthalpy recovery of bulk TPD glasses are analogous to those in volume recovery of TPD thin film shown above. Furthermore, the fictive temperature in Fig. 5d shows an initial plateau before it begins to drop. This initial plateau indicates that the logarithmic aging rate $\beta$ is almost zero at these short times. Similar observations have been reported in molecular,[50] polymeric,[51] and metallic glasses[52], and this feature has also been observed in the volume recovery data shown in Fig. 2. Fig. S12 illustrates the examples for determining $T_f$ of bulk TPD glasses from DSC and FDSC aging measurements at $T_{aging}$=310K. The enthalpy was calculated by integrating the $C_p(T)$ curves.

The volume recovery of TPD thin films and enthalpy recovery of TPD bulk glasses show comparable kinetics when examined in a normalized format. Fig. 6 shows the comparison of $\varphi = \frac{T_{f0}-T_f}{T_{f0}-T_{f\infty}}$ with $T_{f0}$ and $T_{f\infty} = T_{aging}$ denoting the fictive temperature at $t_{aging}$=0 and ∞, for the volume and enthalpy recovery data of TPD glasses determined from SE, DSC, and FDSC measurements. The trend of $\varphi$ with respect to aging time shows very good agreement between SE and DSC measurements; both methods characterized liquid-cooled glasses prepared by cooling at 5 K/min. However, a different dependence of $\varphi$ on $t_{aging}$ is observed in FDSC measurements. This discrepancy originates from the different initial states of the TPD glasses in FDSC measurements, where a much higher quenching rate (1000 K/s) was utilized. In general, supercooled liquids form glasses with higher enthalpy and volume when quenching at higher rates[53]. Therefore, the initial states of TPD glasses in FDSC measurement are expected to have higher energy, compared to those in DSC and SE measurements at the same aging temperature. This is evident when comparing Figs. 2b, 5b, and 5d, where the initial $T_f$ values from FDSC measurements (~352±1 K) are much higher than those from DSC and SE measurements (~332±1 K).

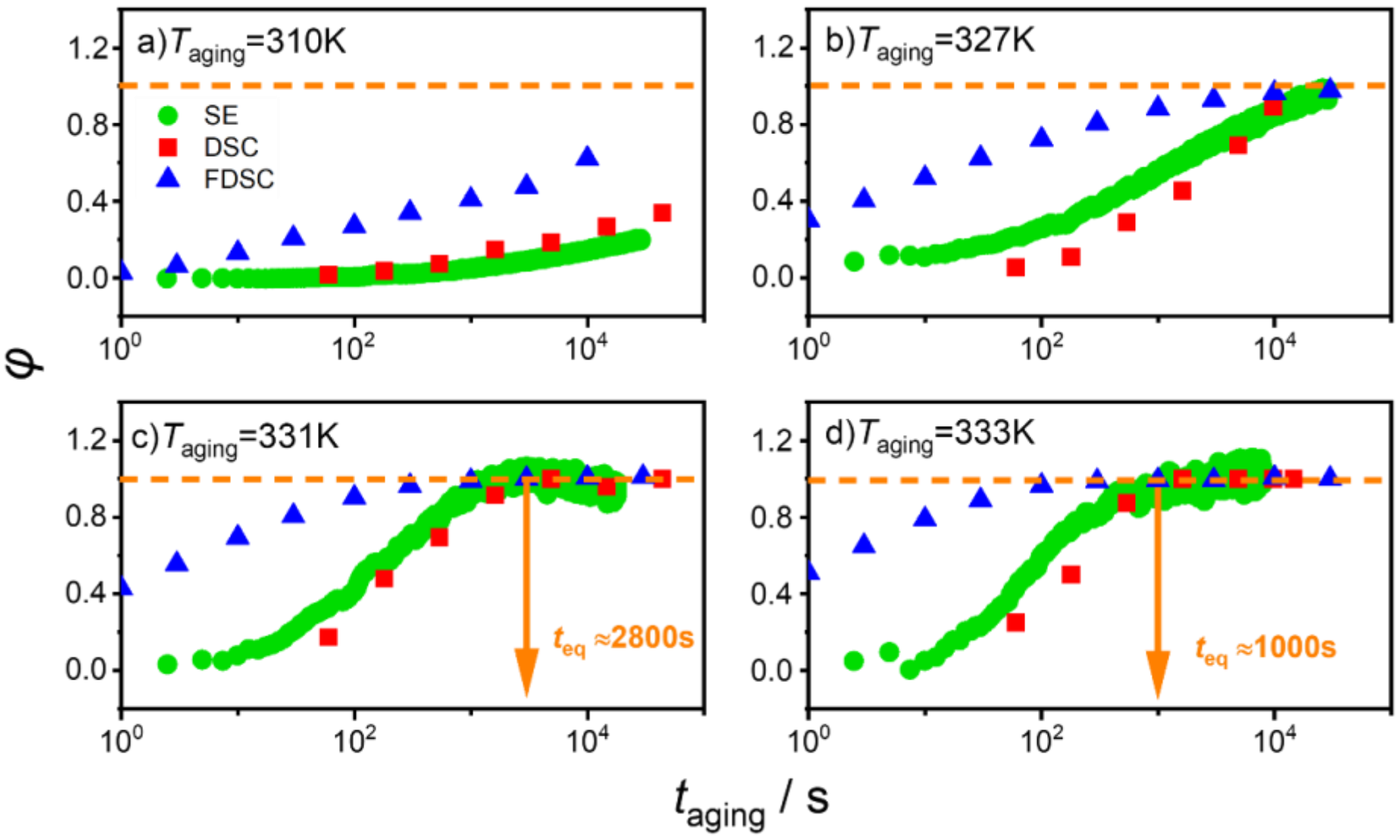

**Fig. 6.** Comparison of physical aging for liquid-cooled TPD glasses as observed by DSC and FDSC (bulk samples) and SE (400 nm films). In each panel, the *y* axis is the normalized recovery (of volume or enthalpy), such that $\varphi$=1 (the dashed organic line) represents the equilibrium state. For the bottom panels, the time required to reach equilibrium $t_{eq}$ is shown to nearly the same for the three techniques.

Despite the difference in the initial state, the TPD glasses in SE, DSC, and FDSC measurements reach the equilibrium state at similar aging times as displayed in Figs. 6c and 6d. In particular, the equilibration times for SE and DSC are almost the same. This is a major result of this work, showing that DSC aging experiments on bulk materials (an experiment available in many laboratories) can provide a reasonable prediction for the aging behavior of thin glass films similar to those used in devices, at least for annealing near $T_g$.

The experiments presented here complement and extend very recent enthalpy relaxation experiments on bulk samples of TPD by Malek and Svoboda[22]. These authors utilized DSC and primarily studied glasses formed by cooling the liquid at different rates. They also performed isothermal aging experiments on bulk samples of TPD at one temperature (318 K) using DSC. While a quantitative comparison between their results and the reported data here would be complex, the qualitative trends in the DSC data from ref.[22] are consistent with those reported here. These authors analyzed their DSC results and previously published ellipsometry data by Zhang and Fakhraai[23] using the Tool–Narayanaswamy–Moynihan (TNM) model. This analysis led to the interesting conclusion that vapor-deposited glasses of TPD exhibit a significantly narrower distribution of relaxation times in comparison to liquid-cooled glasses.

It is useful to consider how the results reported here for the physical aging of vapor-deposited TPD (Figure 3) might be generalized for understanding physical aging in OLED devices. For a typical OLED device, a thin layer (10-60 nm) of an organic semiconductor (or a mixture of organic semiconductors) is vapor-deposited onto either an inorganic substrate or a previously deposited organic semiconductor glass.[54] While we used silicon as the substrate in the experiment shown here (Figure 3), previous work has shown that the stability of PVD glasses is independent of the substrate[55] as long as the substrates are not soft.[42] Regarding film thickness, a recent study showed that the kinetic stability of multilayer PVD glasses of organic semiconductors (with layers as thin as 10 nm) was comparable to that of thick films[56]; this result suggests that the physical aging observed for 400 nm TPD glasses here might be similar to that in actual devices. While many of the layers in actual OLEDs are mixtures, recent work has shown that glass mixtures of co-deposited organic semiconductors can have very high kinetic stability;[57-59] this suggests that the physical aging for vapor-deposited TPD glasses reported here may be indicative of the behavior expected for co-deposited mixtures. As noted above, adjusting the substrate temperature to $0.85T_g$ is expected to further stabilize PVD glasses against physical aging. While the previous work discussed in this paragraph is suggestive, further research is required to directly quantify physical aging of thin layers of organic semiconductors in OLED devices.

## Conclusions

This study provides a comprehensive investigation into the physical aging behavior of glasses of a typical organic semiconductor (TPD) in both thin film and bulk forms, by analyzing film thickness and enthalpy during the aging process. Our results reveal very similar physical aging kinetics (volume relaxation) for 400 nm and 100 nm liquid-cooled TPD films, as evidenced by the highly consistent evolution of $T_f$ with aging time. We interpret this result as consistent with the unchanged glass transition temperature in these films with different thickness. A comparison of the temperature-dependent volume recovery rate for liquid-cooled TPD films to those reported previously for polystyrene films reveals qualitative similarity, but the maximum aging rate for TPD is found somewhat closer to $T_g$ (at 0.96 $T_g$) in comparison to polystyrene (at 0.92 $T_g$). Further work will be required to test whether this is generally true for organic semiconductor glasses.

We also found that the physical aging rate of vapor-deposited TPD glasses is significantly reduced, nearly suppressed by one order of magnitude, compared to that of their liquid-cooled counterparts. This remarkable resistance to physical aging in vapor-deposited organic semiconductors highlights their advantages for use in devices compared to liquid-cooled or solution-processed organic semiconductors. Future research could explore the influence of deposition temperature on the physical aging rate for vapor-deposited TPD glasses, as even lower aging rates are anticipated for deposition near 0.85 $T_g$. TPD glasses deposited at 0.85 $T_g$ have substantial anisotropic molecular orientation, and it would be important to study the evolution of this anisotropy during aging, as molecular orientation is known to strongly influence charge transport properties and thereby device performance.

Finally, our findings demonstrate the important conclusion that volume recovery in both 100 nm and 400 nm TPD films is in good agreement with the enthalpy recovery observed in bulk TPD glasses. We interpret this to indicate a fundamental connection between volume and enthalpy relaxation in organic semiconductor glasses. At a practical level, this observation strongly suggests that studying the physical aging behavior of organic semiconductor glasses in their bulk forms with DSC can provide not only qualitative but also quantitative insights into predicting the time-dependent properties of organic semiconductor thin films, which are widely used in practical applications. This will not only facilitate an improved understanding of structural dynamics in thin films but also aid in the rational design of more stable and efficient organic electronic devices.

## Data availability

The data supporting this article has been included as part of the Supporting Information.

## Conflicts of interest

There are no conflicts of interest to declare.

## Acknowledgements

We thank Sindee L. Simon for insightful discussions. We acknowledge financial support from the National Science Foundation (NSF) through the University of Wisconsin Materials Research Science and Engineering Center (DMR-2309000).

## References

1. A. Dey, A. Singh, D. Das and P. K. Iyer, *Organic semiconductors: a new future of nanodevices and applications*, 2015.
2. F. Li, J. Zhou, J. Zhang and J. Zhao, *Materials*, 2024, **17**, 3362.
3. Y. Wang, Y. Wen, X. Zhuang, S. Liu, L. Zhang and W. Xie, *APL Photonics*, 2024, **9**, 090901.
4. S. Ahmad, *J. Polym. Eng.*, 2014, **34**, 279-338.
5. K. Yoshida, J. Gong, A. L. Kanibolotsky, P. J. Skabara, G. A. Turnbull and I. D. Samuel, *Nature*, 2023, **621**, 746-752.
6. L. Zhou, H.-Y. Xiang, S. Shen, Y.-Q. Li, J.-D. Chen, H.-J. Xie, I. A. Goldthorpe, L.-S. Chen, S.-T. Lee and J.-X. Tang, *ACS nano*, 2014, **8**, 12796-12805.
7. D. Yin, J. Feng, R. Ma, Y.-F. Liu, Y.-L. Zhang, X.-L. Zhang, Y.-G. Bi, Q.-D. Chen and H.-B. Sun, *Nat. Commun.*, 2016, **7**, 11573.
8. M. Micoulaut, *Rep. Prog. Phys.*, 2016, **79**, 066504.
9. I. M. Hodge, *Science*, 1995, **267**, 1945-1947.
10. J. M. HUTCHINSON, *Prog. Polym. Sci.*, 1995, **20**, 703-760.
11. Y. Z. Y. S. L. J. J. d. Christiansen, *Appl. Phys. Lett.*, 2002, **81**, 2983–2985.
12. T. Hecksher, N. B. Olsen, K. Niss and J. C. Dyre, *J. Chem. Phys.*, 2010, **133**, 174514.
13. B. Ruta, E. Pineda and Z. Evenson, *J. Phys. Condens. Matter*, 2017, **29**, 503002.
14. G. B. McKenna, Y. Leterrier and C. R. Schultheisz, *Polym. Eng. Sci.*, 1995, **35**, 403-410.
15. S. L. Simon, D. J. Plazek, J. W. Sobieski and E. T. McGregor, *J. Polym. Sci., Part B:Polym. Phys.*, 1997, **35**, 929-936.
16. P. Pan, B. Zhu and Y. Inoue, *Macromolecules*, 2007, **40**, 9664-9671.
17. J. M. Cowie and R. Ferguson, *Macromolecules*, 1989, **22**, 2307-2312.
18. V. A. Soloukhin, J. C. Brokken-Zijp, O. L. van Asselen and G. de With, *Macromolecules*, 2003, **36**, 7585-7597.
19. C. T. Powell, H. Xi, Y. Sun, E. Gunn, Y. Chen, M. Ediger and L. Yu, *J. Phys. Chem. B*, 2015, **119**, 10124-10130.
20. P. Smith, P. Gerroir, S. Xie, A. Hor, Z. Popovic and M. Hair, *Langmuir*, 1998, **14**, 5946-5950.

21. M. Azrain, M. Mansor, S. Fadzullah, G. Omar, D. Sivakumar, L. Lim and M. Nordin, *Synth. Met.*, 2018, **235**, 160-175.
22. J. Málek and R. Svoboda, *J. Chem. Phys.*, 2024, **161**, 074507.
23. Y. Zhang and Z. Fakhraai, *Phys. Rev. Lett.*, 2017, **118**, 066101.
24. T. Hecksher, N. B. Olsen and J. C. Dyre, *Proc. Natl. Acad. Sci. U.S.A.*, 2019, **116**, 16736-16741.
25. E. F. Oleinik, *Polym. J.*, 1987, **19**, 105-117.
26. K. Arabeche, L. Delbreilh and E. Baer, *J. Polym. Res.*, 2021, **28**, 431.
27. J. E. Pye, K. A. Rohald, E. A. Baker and C. B. Roth, *Macromolecules*, 2010, **43**, 8296-8303.
28. D. Cangialosi, *J. Polym. Sci.*, 2024, 1952-1974.
29. D. M. Walters, R. Richert and M. D. Ediger, *J. Chem. Phys.*, 2015, **142**, 134504.
30. S. S. Dalal and M. D. Ediger, *J. Phys. Chem. Lett.*, 2012, **3**, 1229-1233.
31. S. S. Dalal, Z. Fakhraai and M. D. Ediger, *J. Phys. Chem. B*, 2013, **117**, 15415-15425.
32. E. Leon-Gutierrez, A. Sepúlveda, G. Garcia, M. T. Clavaguera-Mora and J. Rodríguez-Viejo, *Phys. Chem. Chem. Phys.*, 2010, **12**, 14693-14698.
33. S. L. L. Ramos, M. Oguni, K. Ishii and H. Nakayama, *J. Phys. Chem. B*, 2011, **115**, 14327-14332.
34. S. F. Swallen, K. L. Kearns, M. K. Mapes, Y. S. Kim, R. J. McMahon, M. D. Ediger, T. Wu, L. Yu and S. Satija, *Science*, 2007, **315**, 353-356.
35. K. L. Kearns, P. Krzyskowski and Z. Devereaux, *J. Chem. Phys.*, 2017, **146**, 203328.
36. S. S. Dalal, D. M. Walters, I. Lyubimov, J. J. de Pablo and M. D. Ediger, *Proc. Natl. Acad. Sci. U.S.A.*, 2015, **112**, 4227-4232.
37. K. Bagchi, M. E. Fiori, C. Bishop, M. Toney and M. Ediger, *J. Phys. Chem. B*, 2020, **125**, 461-466.
38. C. Rodriguez-Tinoco, M. Gonzalez-Silveira, M. A. Ramos and J. Rodriguez-Viejo, *La Rivista del Nuovo Cimento*, 2022, **45**, 325-406.
39. P. Luo and Z. Fakhraai, *Annu. Rev. Phys. Chem.*, 2023, **74**, 361-389.
40. H. Mu, W. Li, R. Jones, A. Steckl and D. Klotzkin, *J. Lumin.*, 2007, **126**, 225-229.
41. S. S. Swayamprabha, M. R. Nagar, R. A. K. Yadav, S. Gull, D. K. Dubey and J.-H. Jou, *J. Mater. Chem. C*, 2019, **7**, 7144-7158.
42. P. Luo, S. E. Wolf, S. Govind, R. B. Stephens, D. H. Kim, C. Y. Chen, T. Nguyen, P. Wąsik, M. Zhernenkov and B. Mcclimon, *Nat. Mater.*, 2024, **23**, 688-694.
43. Y. Jin, A. Zhang, S. E. Wolf, S. Govind, A. R. Moore, M. Zhernenkov, G. Freychet, A. Arabi Shamsabadi and Z. Fakhraai, *Proc. Natl. Acad. Sci. U.S.A.*, 2021, **118**, e2100738118.
44. E. A. Baker, P. Rittigstein, J. M. Torkelson and C. B. Roth, *J. Polym. Sci., Part B:Polym. Phys.*, 2009, **47**, 2509-2519.
45. K. Dalnoki-Veress, J. Forrest, C. Murray, C. Gigault and J. Dutcher, *Phys. Rev. E*, 2001, **63**, 031801.
46. Y. Zhang, C. N. Woods, M. Alvarez, Y. Jin, R. A. Riggleman and Z. Fakhraai, *J. Chem. Phys.*, 2018, **149**, 184902.
47. V. M. Boucher, C. Daniele, A. a. Angel and C. Juan, *Macromolecules*, 2012, **45**, 5296-5306.
48. M. Moratalla, M. Rodríguez-López, C. Rodríguez-Tinoco, J. Rodríguez-Viejo, R. J. Jiménez-Riobóo and M. A. Ramos, *Commun. Phys.*, 2023, **6**, 274.
49. M. Shibata, Y. Sakai and D. Yokoyama, *J. Mater. Chem. C*, 2015, **3**, 11178-11191.
50. V. Di Lisio, V.-M. Stavropoulou and D. Cangialosi, *J. Chem. Phys.*, 2023, **159**, 064505.
51. Y. P. Koh, S. Gao and S. L. Simon, *Polymer*, 2016, **96**, 182-187.
52. J. Q. Wang, Y. Shen, J. H. Perepezko and M. D. Ediger, *Acta Mater*, 2016, **104**, 25-32.
53. M. D. Ediger, C. A. Angell and S. R. Nagel, *J. Phys. Chem.*, 1996, **100**, 13200-13212.

54. J. Ràfols-Ribé, P.-A. Will, C. Hänisch, M. Gonzalez-Silveira, S. Lenk, J. Rodríguez-Viejo and S. Reineke, *Sci. Adv.*, 2018, **4**, eaar8332.
55. K. L. Kearns, S. F. Swallen, M. Ediger, T. Wu and L. Yu, *J. Chem. Phys.*, 2007, **127**, 154702.
56. M. E. Fiori, K. Bagchi, M. F. Toney and M. D. Ediger, *Proc. Natl. Acad. Sci. U.S.A.*, 2021, **118**, e2111988118.
57. Y. Lee, S. Cheng and M. Ediger, *J. Phys. Chem. Lett.*, 2024, **15**, 8085-8092.
58. S. Cheng, Y. Lee, J. Yu, L. Yu and M. D. Ediger, *Chem. Mater.*, 2024, **36**, 3205-3214.
59. S. Cheng, Y. Lee, J. Yu, L. Yu and M. D. Ediger, *J. Phys. Chem. Lett.*, 2023, **14**, 4297-4303.

# Supporting information for

## Physical aging of glasses of an organic semiconductor

Shinian Cheng[1*], Kritika Jha[2], Zijian Wang[3], Juliana B. Lugo[2], Hayley Kositzke[1],
John Perepezko[3], Zahra Fakhraai[2], Mark D. Ediger[1]

[1]Department of Chemistry, University of Wisconsin-Madison, Madison, Wisconsin 53706, United States
[2]Department of Chemistry, University of Pennsylvania, Philadelphia, Pennsylvania 19104, United States
[3]Department of Materials Science and Engineering, University of Wisconsin-Madison, Madison, Wisconsin 53706, United States

*Corresponding Author: chengshinian@gmail.com;

This supporting information file includes:

**1. Temperature profiles for aging experiments**

**2. Representative fitting parameters of the isotropic Cauchy model for the liquid-cooled TPD film**

**3. Representative fitting parameters of the anisotropic Cauchy model for vapor-deposited TPD film**

**4. Refractive index for TPD films in glassy and supercooled liquid states**

**5. Determination of $T_f$ of TPD films from SE measurements**

**6. Physical aging rate of liquid-cooled TPD films**

**7. DSC and FDSC thermograms of bulk TPD glasses**

**8. Examples for determining $T_f$ of bulk TPD glasses from DSC and FDSC measurements**

## 1. Temperature profiles for aging experiments

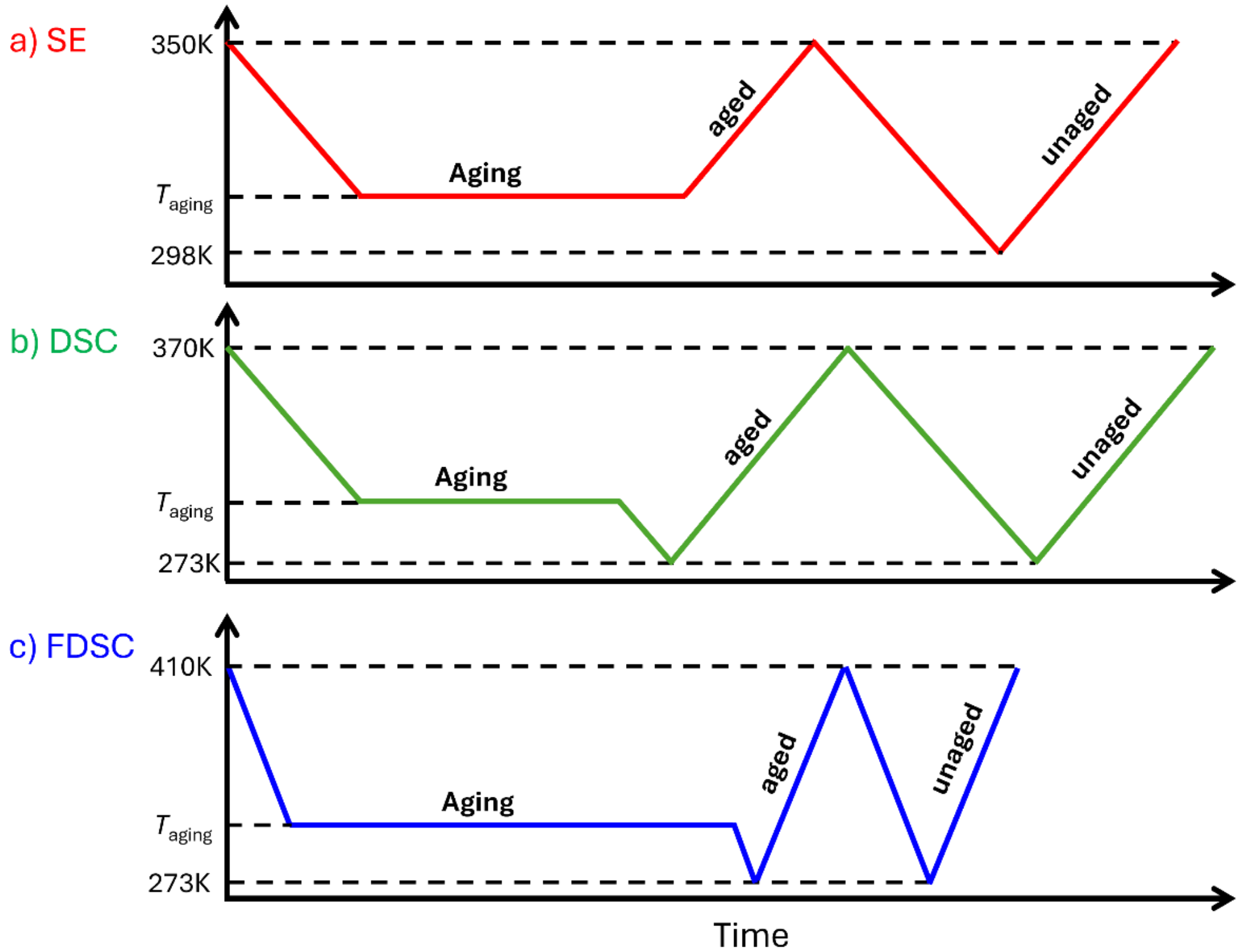

**Fig. S1:** The temperature profiles of aging experiments on liquid-cooled TPD glasses for a) SE, b) DSC, and c) FDSC measurements.

## 2. Representative fitting parameters of the isotropic Cauchy model for the liquid-cooled TPD film

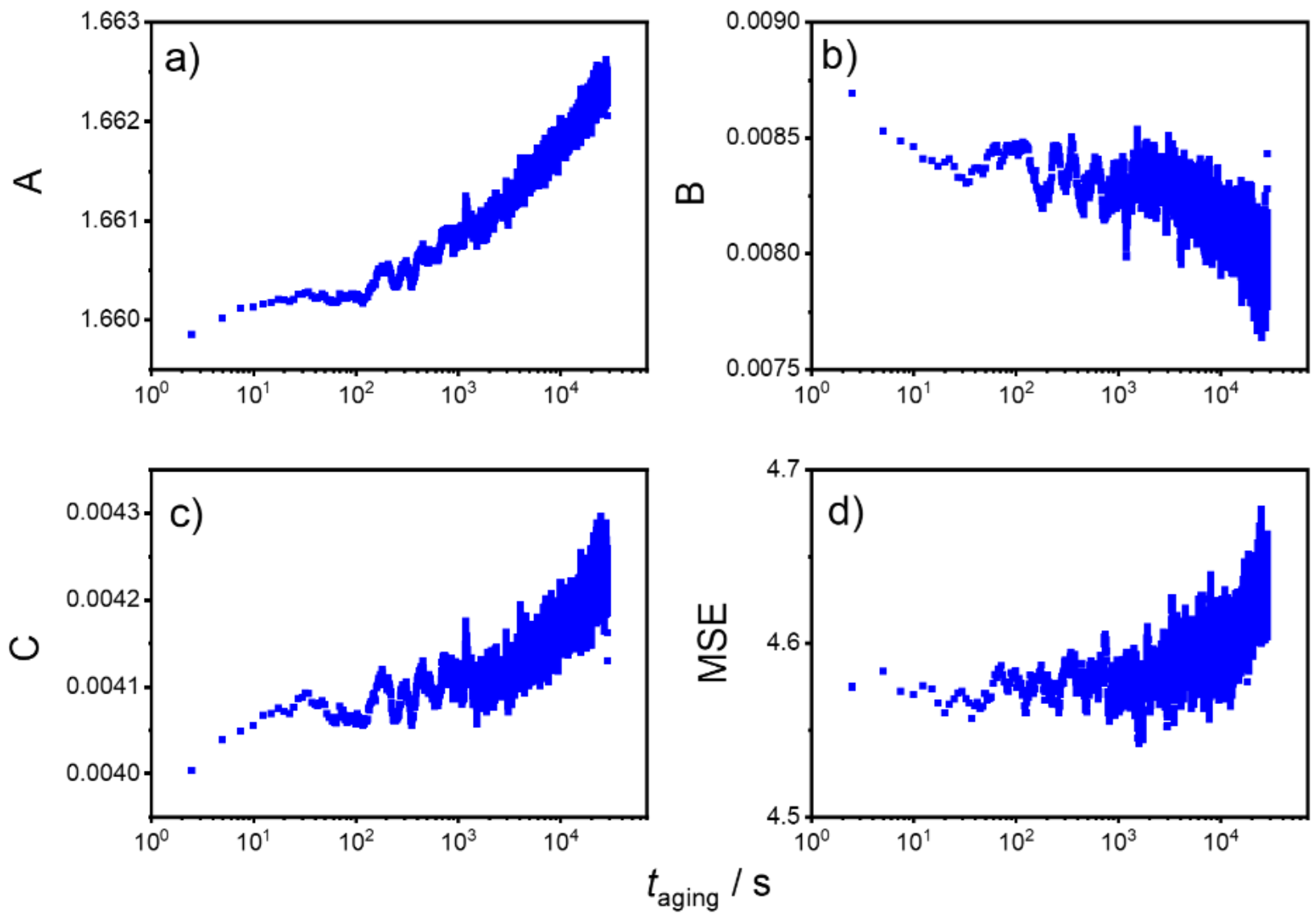

**Fig. S2:** Representative isotropic Cauchy model parameters for a 400 nm liquid-cooled TPD film as a function of aging time at $T_{aging}$=318K. **d)** The obtained Mean Squared Error (MSE) between the model and experimental data.

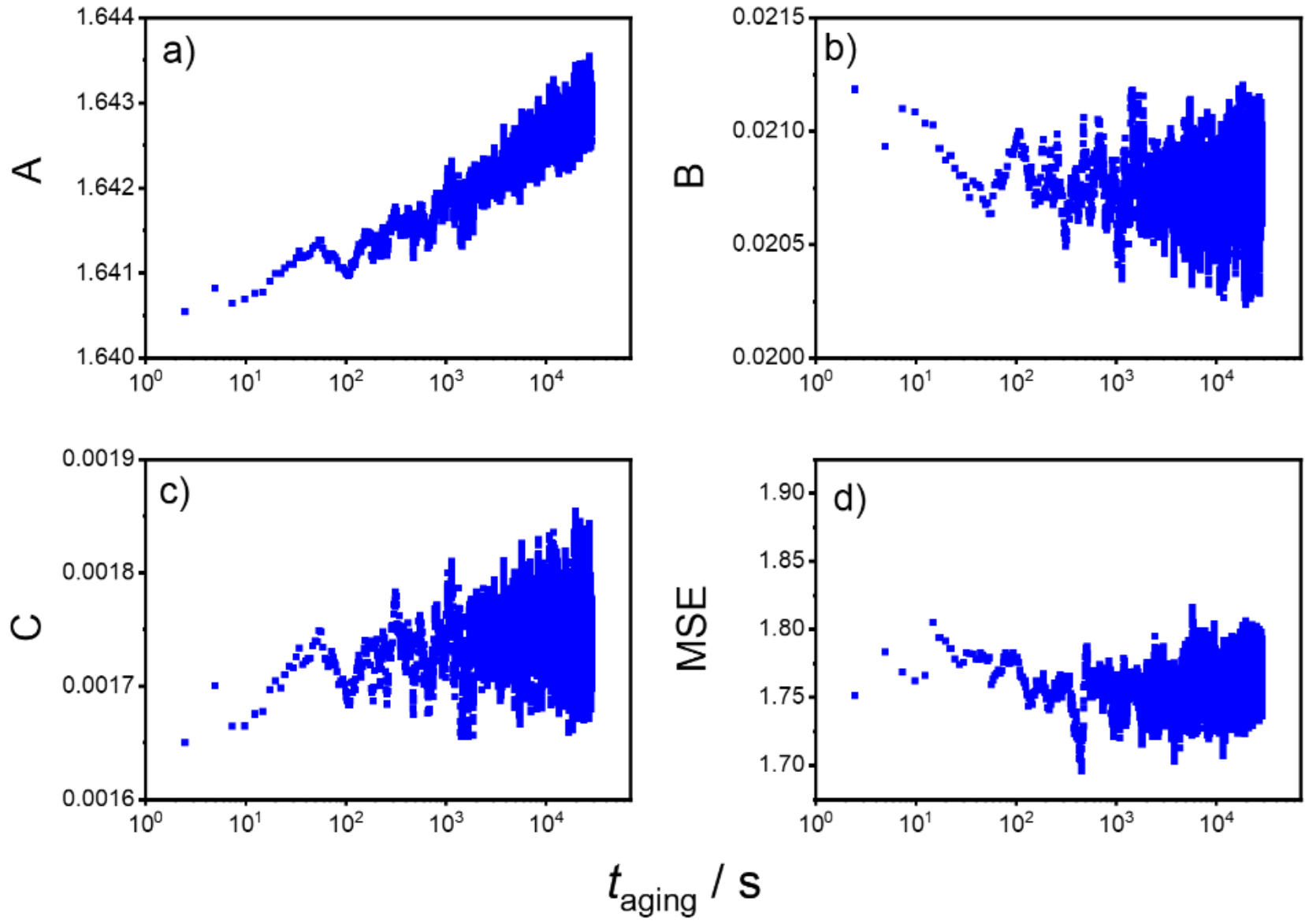

**Fig. S3:** Representative isotropic Cauchy model parameters for a 100 nm liquid-cooled TPD film as a function of aging time at $T_{aging}$=318K. **d)** The obtained Mean Squared Error (MSE) between the model and experimental data

## 3. Representative fitting parameters of the anisotropic Cauchy model for vapor-deposited TPD film

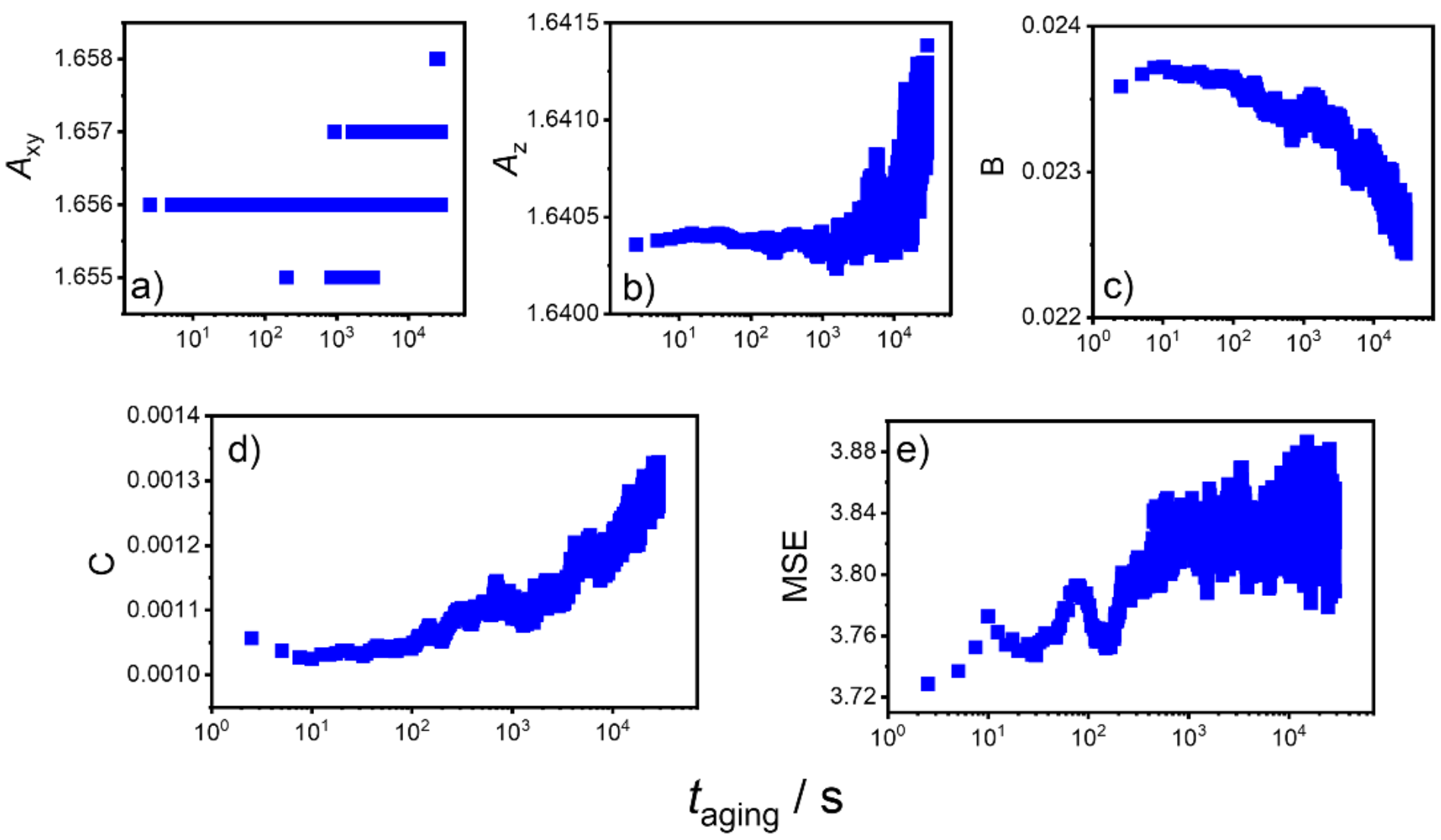

**Fig. S4:** Representative anisotropic Cauchy model parameters for a 400 nm vapor-deposited TPD film as a function of aging time at $T_{aging}$=318K. **d)** The obtained Mean Squared Error (MSE) between the model and experimental data.

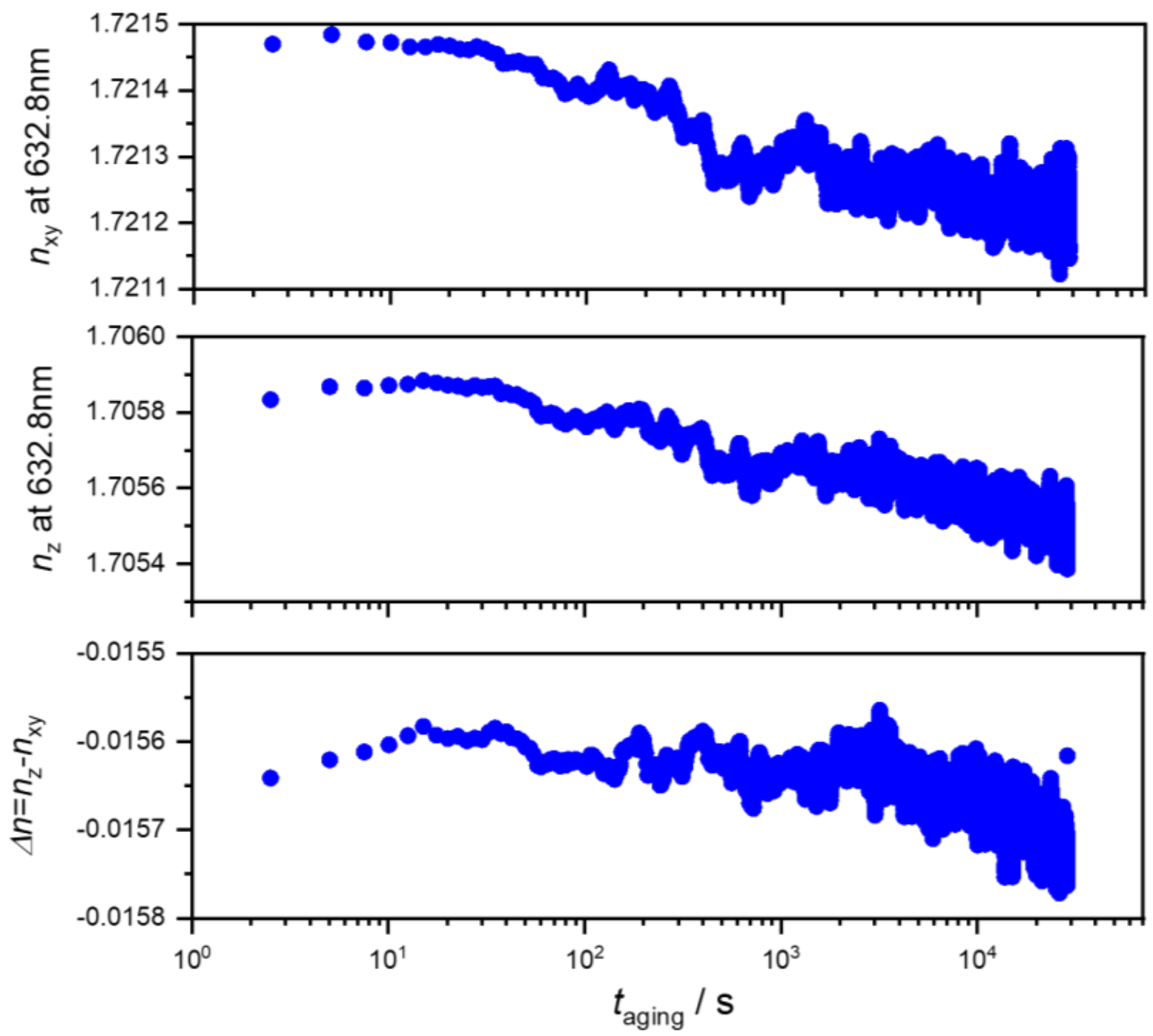

**Fig. S5:** The evolution of refractive indices **a)** $n_{xy}$ and **b)** $n_z$ at 632.8 nm with aging time for a 400 nm vapor-deposited TPD film at $T_{aging}$=318K. **c)** The determined birefringence, $\Delta n = n_z - n_{xy}$, at 632.8 nm as a function of aging time for vapor-deposited TPD film at $T_{aging}$=318K.

## 4. Refractive index for TPD films in glassy and supercooled liquid states

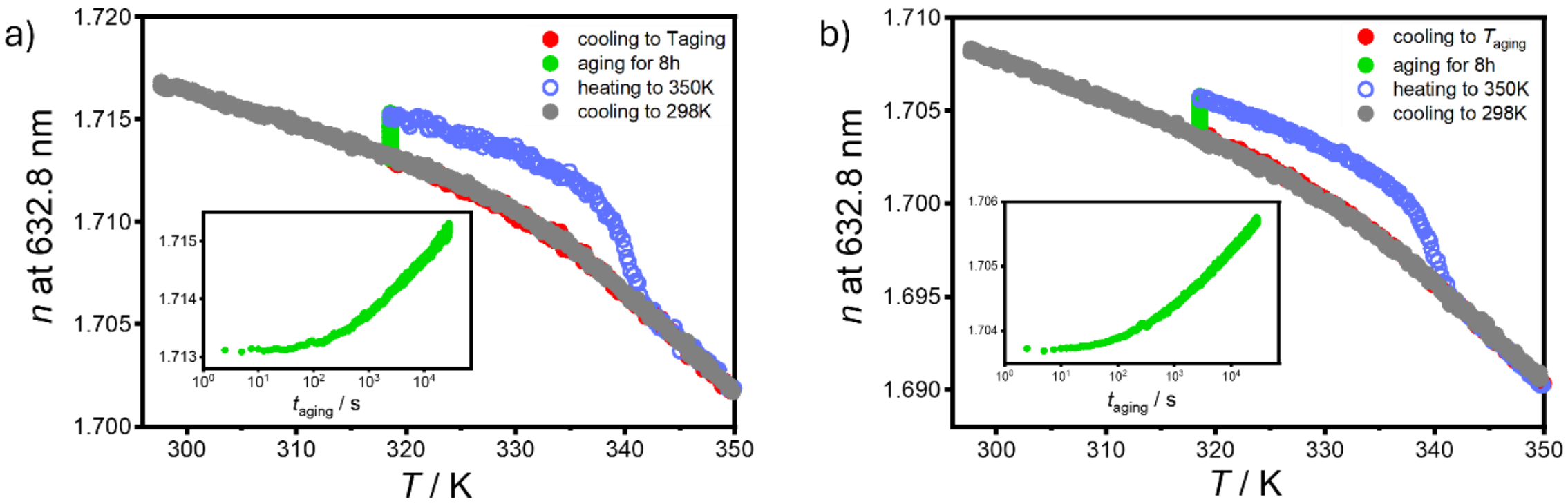

**Fig. S6:** The refractive index $n$ at 632.8 nm as a function of temperature during an aging experiment at $T_{\mathrm{aging}}$=318K for **a)** 400 nm and **b)** 100 nm TPD glasses and supercooled liquid.

## 5. Determination of $T_{\mathrm{f}}$ of TPD films from SE measurements

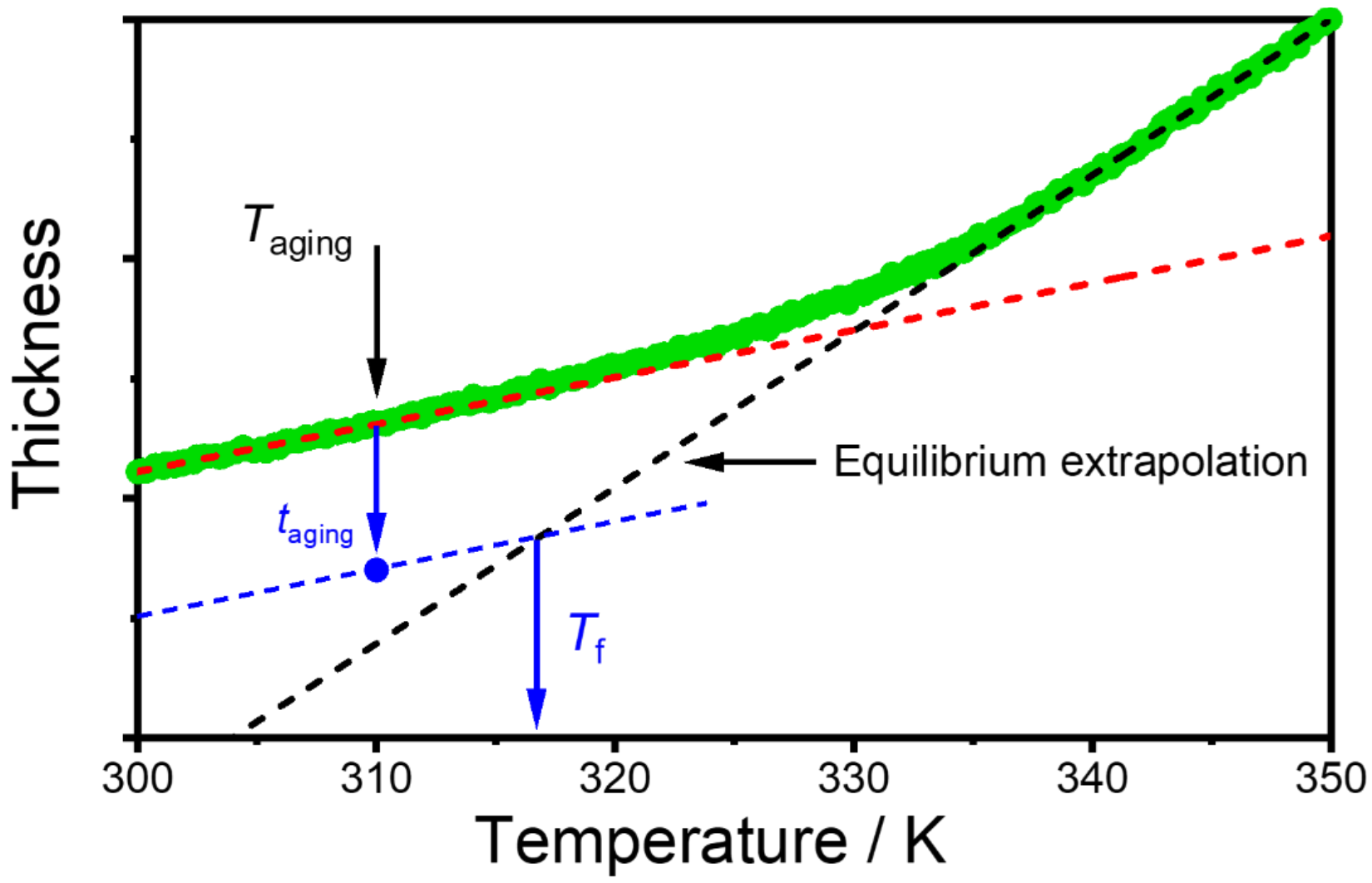

**Fig. S7:** Illustration of the method used to calculate $T_{\mathrm{f}}$ of aged TPD glasses at a given aging temperature $T_{\mathrm{aging}}$. The blue dot represents the glass film thickness at $T_{\mathrm{aging}}$ after aging for $t_{\mathrm{aging}}$. The dashed blue line indicates the predicted evolution of the film thickness with temperature of the aged glass, characterized by the expansion coefficient $G$ determined from Eq. 1. The function of this blue line can be described as $h$-$h_{\mathrm{aging}}$=$G$×($T$-$T_{\mathrm{aging}}$), where $h_{\mathrm{aging}}$ is the glass film thickness after aging for $t_{\mathrm{aging}}$ at $T_{\mathrm{aging}}$. The black line represents the extrapolated thickness of the supercooled liquid with expansion coefficient $M$ determined from Eq. 1. According to the definition of $T_{\mathrm{f}}$, its value corresponds to the intersection of the blue and black lines. Therefore, $h_{\mathrm{aging}}$ as a function of $t_{\mathrm{aging}}$ at a specific $T_{\mathrm{aging}}$ determined from SE measurements is converted into $T_{\mathrm{f}}$ as a function of $t_{\mathrm{aging}}$.

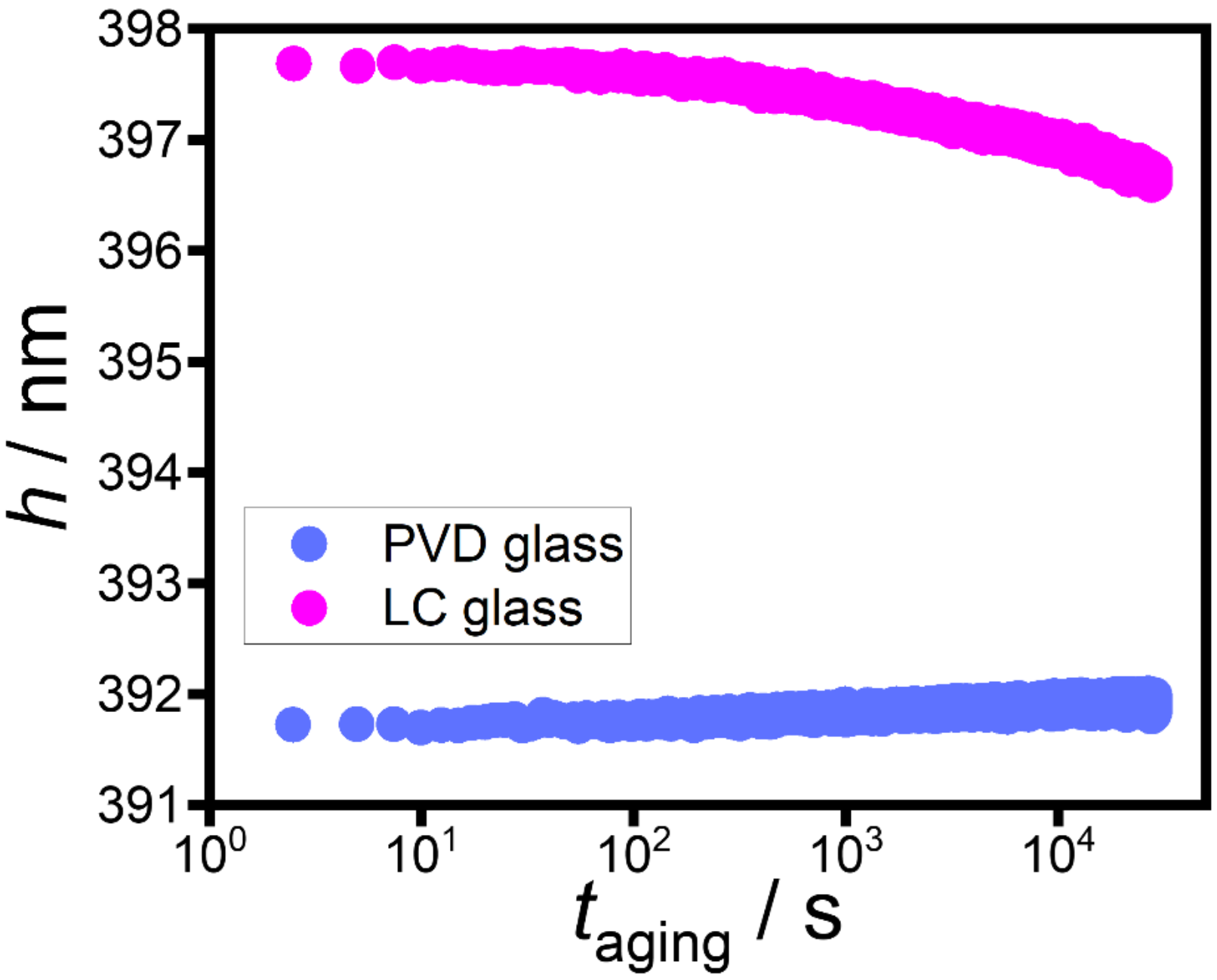

**Fig. S8:** The thickness as a function of aging time for vapor-deposited (blue) and liquid-cooled (pink) TPD thin films at $T_{aging}$=318 K.

## 6. Physical aging rate of liquid-cooled TPD films

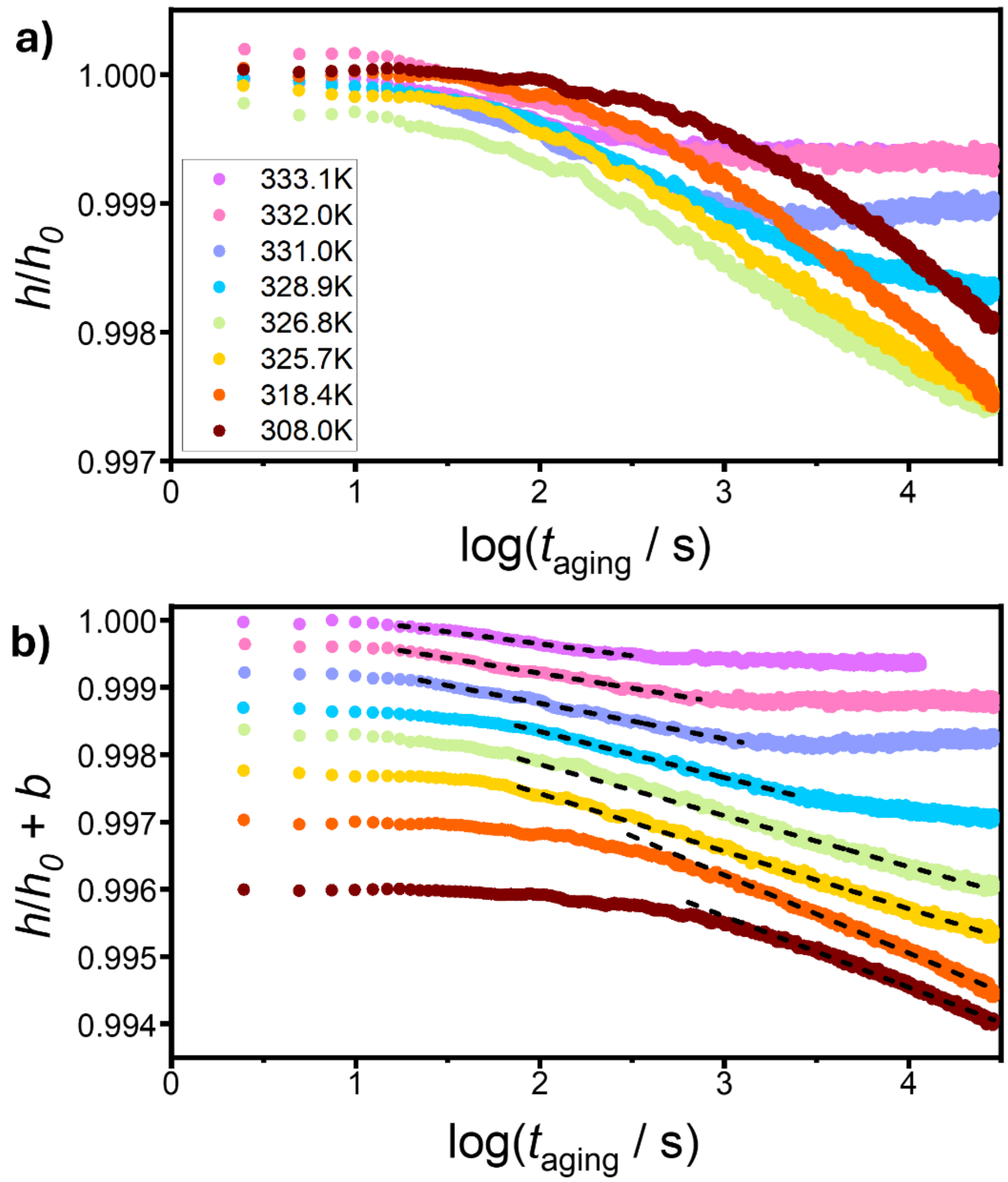

**Fig. S9: a)** Normalized thickness is plotted as a function of log($t_{aging}$) of liquid-cooled TPD films at various aging temperatures. $h_0$ is the film thickness at $t_{aging}$=0 s; **b)** Illustration of the linear fits (dashed lines) to obtain physical aging rates of liquid-cooled TPD films at various aging temperatures. The *Y* axis is shifted by an amplitude of *b* for the presentation purposes.

## 7. DSC and FDSC thermograms of bulk TPD glasses

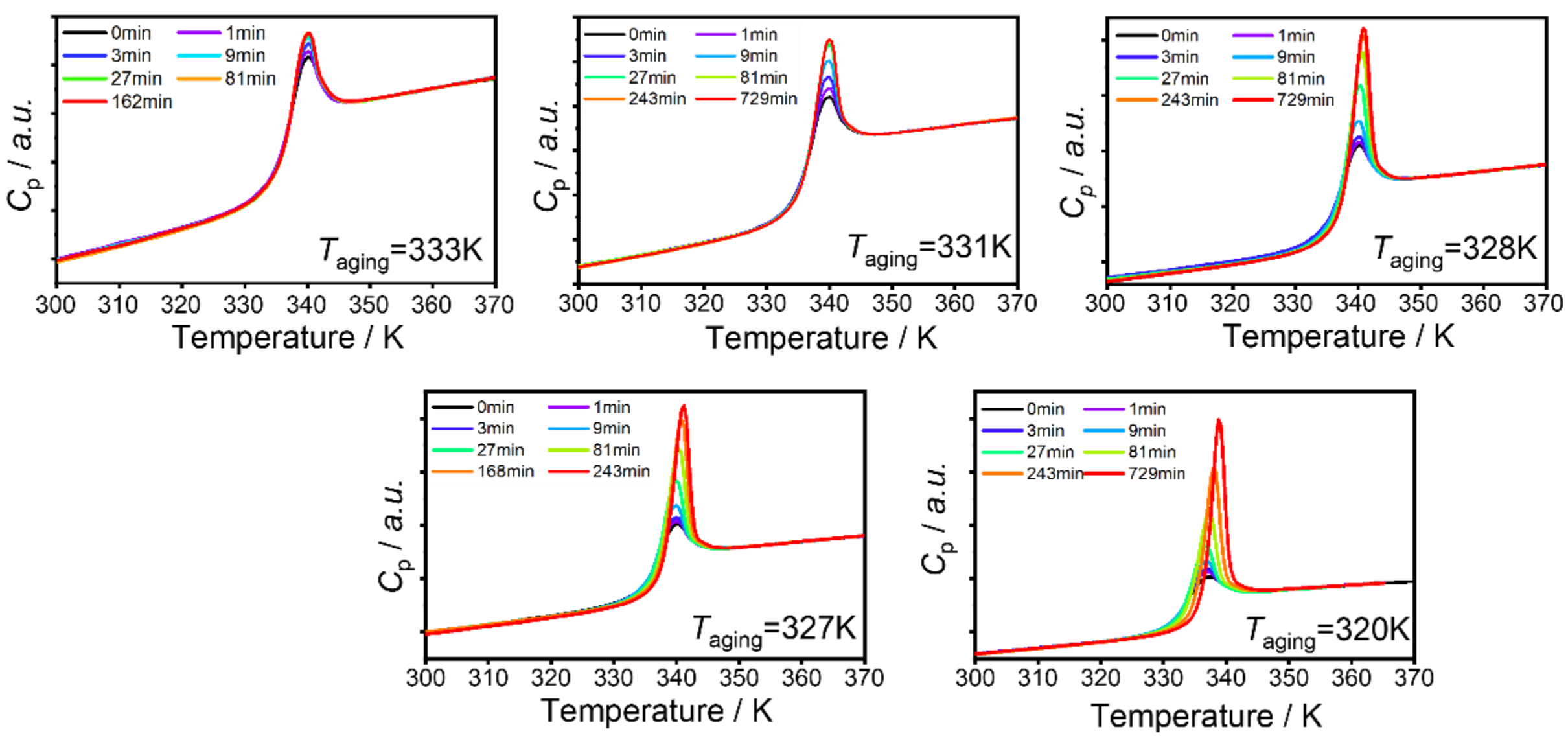

**Fig. S10:** DSC thermograms of bulk TPD glasses after aging at 333 K, 331K, 328K, 327K, and 320K, for the indicated aging times.

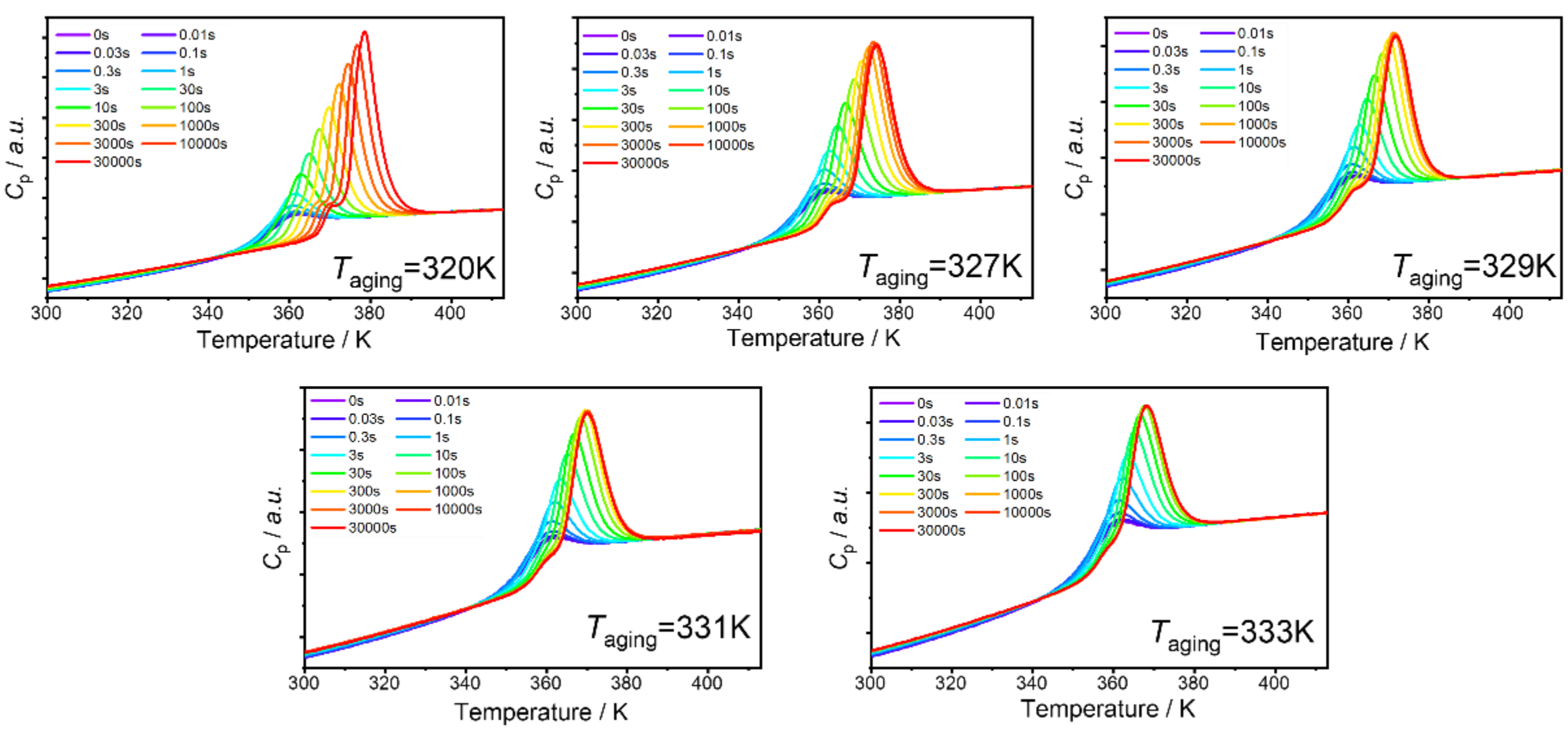

**Fig. S11:** FDSC thermograms of bulk TPD glasses after aging at 320 K, 327K, 329K, 331K, and 333K for the indicated aging times.

## 8. Examples for determining $T_f$ of bulk TPD glasses from DSC and FDSC measurements

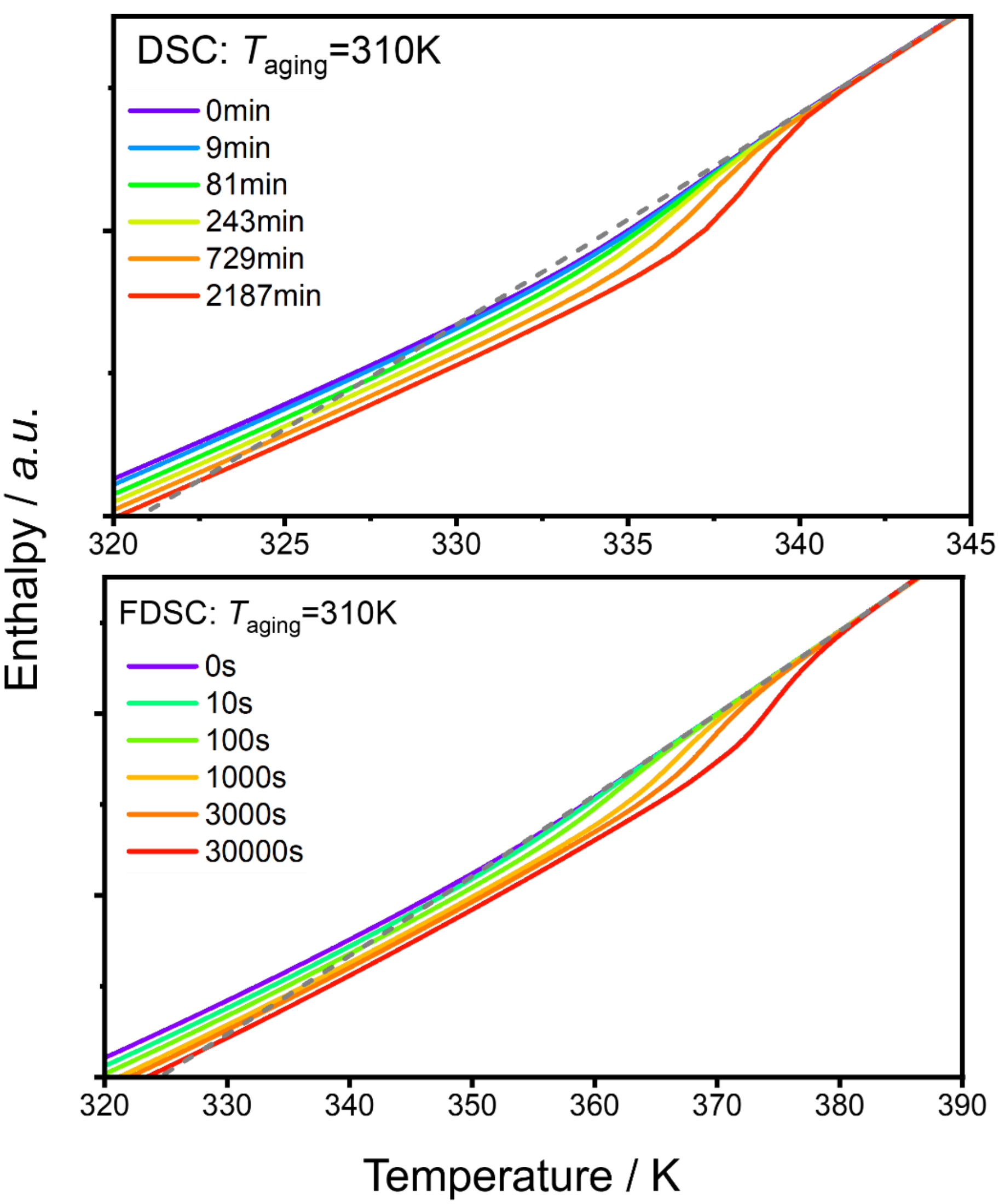

**Fig. S12:** Enthalpy as a function of temperature for bulk TPD glasses aged at 310K for various aging times. The dashed lines present the extrapolated enthalpy of supercooled liquid of TPD using a second-order polynomial function: $\Delta H$=a+b×$T$+c×$T^2$. The $T_f$ values are determined as the intersections of the dashed gray and the solid lines, which present the temperature-dependent enthalpy of the TPD glasses.